\documentclass[compsoc,conference,a4paper,10pt,times]{IEEEtran} % this is for EuroS&P / SP

\usepackage[dvipsnames]{xcolor}

\usepackage[colorlinks=true,linkcolor=blue,breaklinks=True,citecolor=brown,urlcolor=blue]{hyperref}

\usepackage{dingbat}
\usepackage[utf8]{inputenc}
\usepackage{multirow}
\usepackage{booktabs}
\usepackage{rotating}
\usepackage{adjustbox}
\usepackage{threeparttable}
\usepackage{longtable}
\usepackage{multirow}
\usepackage{booktabs}
\usepackage{rotating}
\usepackage[linesnumbered,algo2e,ruled]{algorithm2e}
\usepackage{multirow}
\usepackage{booktabs,subcaption,dcolumn}
\usepackage{tikz}
\usepackage{enumitem}
\setlist[itemize]{noitemsep, topsep=0pt}
\usepackage{tabularx}
\usepackage[normalem]{ulem}
\usepackage{todonotes}
\usepackage{threeparttable}

\usepackage{graphicx}
\usepackage{url}
\usepackage{stfloats}
\usepackage{placeins}
\usepackage{placeins}
\usepackage{comment}

\usepackage{pifont}% http://ctan.org/pkg/pifont
\usepackage{svg}
\usepackage{soul}
\usepackage[font=footnotesize]{caption}
\usepackage[noadjust]{cite}

\usepackage{amsmath}
\usepackage{amsfonts, amssymb}
\usepackage{colortbl}
\usepackage{tablefootnote}
\usepackage[most]{tcolorbox}
\usepackage[sort&compress, numbers]{natbib}

\usepackage{textcmds}
\AtBeginDocument{%
  \providecommand\BibTeX{{%
    \normalfont B\kern-0.5em{\scshape i\kern-0.25em b}\kern-0.8em\TeX}}}

\newcolumntype{?}{!{\vrule width 1.5pt}}

\newtcolorbox{cooltextbox}[1][]{%
    colback=black!5,
    colframe=black!5,
    notitle,
    sharp corners,
    borderline west={0pt}{0pt}{red!80!black},
    enhanced,
    breakable,
    left=0pt,
    right=0pt,
    top=0pt,
    bottom=0pt
    }

\newcommand\revision[1]{%
  \bgroup
  \hskip0pt\color{blue!80!black}%
  #1%
  \egroup
}

\begin{document}

% \title{LA-DW: Longitudinal Analysis for Dark Web}
\title{Topical Shifts in the Dark Web: A Longitudinal Analysis of Content from the Cybercrime Ecosystem}
% \titlerunning{Use this for LNCS if the title is too long}

\author{\IEEEauthorblockN{ {
Roy Ricaldi}}
\IEEEauthorblockA{\textit{Department of Computer Science} \\
\textit{Eindhoven University of Technology}\\
Eindhoven, Netherlands \\
r.j.ricaldi.saavedra@tue.nl}
\and
\IEEEauthorblockN{Maximilian Schäfer}
\IEEEauthorblockA{\textit{Department of Information Systems} \\
\textit{University of Liechtenstein}\\
Vaduz, Liechtenstein \\
maximilian.schaefer@uni.li}
\and
\IEEEauthorblockN{Philipp Zech}
\IEEEauthorblockA{\textit{Department of Computer Science} \\
\textit{University of Innsbruck}\\
Innsbruck, Austria \\
philipp.zech@uibk.ac.at}
\and
\IEEEauthorblockN{
Luca Allodi}
\IEEEauthorblockA{\textit{Department of Computer Science} \\
\textit{Eindhoven University of Technology}\\
Eindhoven, Netherlands \\
l.allodi@tue.nl}
\and
\IEEEauthorblockN{Raffaela Groner}
\IEEEauthorblockA{\textit{Department of Computer Science} \\
\textit{Chalmers University of Technology}\\
Gothenburg, Sweden \\
raffaela@chalmers.se}
\and
\IEEEauthorblockN{Irdin Pekaric}
\IEEEauthorblockA{\textit{Department of Information Systems} \\
\textit{University of Liechtenstein}\\
Vaduz, Liechtenstein \\
irdin.pekaric@uni.li}
}

\pagestyle{plain}
\maketitle

\begin{abstract}
% The dark web serves as a valuable source of information for cyber threat intelligence (CTI). Yet, despite over two decades of existence, it remains largely underexplored. Recent studies mainly focus on the extraction of operational CTI. However, strategic CTI, which requires a more comprehensive and long-term analysis of dark web dynamics, has been overlooked. This study addresses this gap by conducting a longitudinal analysis of darknet websites, employing a combination of Machine Learning (ML) and Natural Language Processing (NLP) techniques. Our research examines (n=?) darknet websites by targeting those with at least four snapshots over time to track their evolution. By analyzing content update frequencies, topic emergence and disappearance, external URLs changes, and the interplay between related URLs and topic shifts, we aim to provide a comprehensive understanding of the cybercrime ecosystem in the dark web. 

The dark web hosts a dynamic ecosystem of cybercrime forums and marketplaces that adapt to law enforcement pressure, technological change, and economic incentives. Prior research has extracted cyber threat intelligence from these platforms using static snapshots, with limited attention to how discussions evolve over time. 

In this study, we conduct a longitudinal analysis of 25,065 websites in the dark web using 11,403,638 HTML snapshots (approximately 1245.38 GB) collected over six years. We develop a longitudinal topic-modeling framework combining domain-specific embeddings, density-based clustering and temporal aggregation to measure topic prevalence and lifecycle at the website level. Our analysis identifies 55 thematic clusters. We find that $\approx75\%$ of total discussion volume is concentrated in a small set of persistent core topics, while short-lived themes account for $\approx3\%$ of activity. The median topic lifespan is 75 months, indicating gradual thematic evolution rather than abrupt replacement.

\end{abstract}


\section{Introduction}
\label{sec:introduction}

\noindent The dark web constitutes a hidden segment of the internet that enables anonymous communication and hosting of services through specialized routing protocols such as Tor. While it provides privacy protections for journalists, activists, and whistleblowers, it also serves as a major infrastructure for cybercrime. Prior measurements estimate that 57\% of dark web content is associated with illicit activity \cite{essien2020relevance}, and Tor Metrics data indicate that daily users have doubled since 2022, reaching a peak of 7 million users \cite{tormetrics}. This growth highlights the continued importance of the dark web as a space where anonymity, financial motives, and technology intersect.

Despite more than two decades of existence, the dark web remains only partially understood, particularly with regard to how cybercrime-related content evolves over time \cite{9842561,11129557}. Research has largely approached dark web platforms through Cyber Threat Intelligence (CTI), focusing on extracting actionable signals such as indicators of compromise, exploit discussions, or marketplace listings. Early work relied on dictionary-based or fuzzing techniques, often suffering from limited recall or high false positives \cite{robertson2017darkweb,7745435}. More recent studies apply machine learning and natural language processing to improve detection and classification of threat-relevant information \cite{11315706}.

However, most existing studies treat dark web data as static snapshots \cite{paoli2017behind,ball2021data} or focus on a limited set of marketplaces rather than the broader cybercrime ecosystem \cite{hoheisel2026assessing,decary2017police}. While useful for operational intelligence, these approaches provide limited insight into how cybercrime-related discussions evolve. This is a fundamental limitation in complex threat environments, where security analysis must move beyond static observations and adjust to a continuously evolving attacker ecosystem \cite{felderer2017research, pekaric2025saftgt}. This demonstrates the need for similarly changing perspectives in CTI.
%In particular, it remains unclear whether observed topic shifts reflect lasting structural changes or temporary fluctuations.
Dark web platforms are highly dynamic: platforms shut down, rebrand, or migrate; forums fragment or consolidate; and marketplaces adapt to law enforcement pressure, technological developments, and user demand. Yet, how topics and themese discussed and of relevance in these communities and websites emerge, persist, and disappear over time has not been systematically studied longitudinally. Without this perspective, it is difficult to distinguish stable core activities from transient trends.

This study addresses this gap by reporting on a longitudinal analysis of dark web forums and marketplaces using automated methods for natural language processing and ML. Relying on repeated HTML snapshots spanning from 2020 to 2026, we analyze how (criminal) content evolves over time, capturing how topics emerge, persist, and decline.

\smallskip
\noindent \textsc{\textbf{Contributions.}}
We make the following contributions:
%\vspace{-7mm}
\begin{itemize}[leftmargin=*,noitemsep,topsep=0pt]
\item We propose a longitudinal measurement framework for analyzing forum and marketplace websites in the dark web using HTML snapshots.

\item We develop a topic\footnote{We consider specific content categories (e.g., botnets, ransomware, forged document services) as topics rather than very broad umbrella themes such as drugs or weapons.} discovery pipeline that integrates DARK-BERT embeddings, density-based clustering, and probabilistic topic assignment to model the temporal dynamics of discussion themes.

\item Using a dataset of over 11 million snapshots from more than 25 thousand dark web websites collected over six years, we provide the first large-scale longitudinal analysis of topic prevalence and turnover in economically motivated cybercrime communities.

\end{itemize}

\vspace{2mm}

We release our resources to support software sustainability: \url{https://github.com/irdin-pekaric/WACCO2026}. 
\section{Background and Related Work}
\label{sec:related}

\noindent We provide background information on cybercrime ecosystems and the dark web (§\ref{ssec:backgroundDarknet}). In addition, we discuss related work on methods that analyze websites on the dark web (§\ref{ssec:related}) and pinpoint the research gap (§\ref{ssec:research_gap}).

\subsection{Cybercrime in the Dark Web} \label{ssec:backgroundDarknet}
\noindent
For this study, \texttt{cybercrime} refers to illicit activities conducted through digital systems and primarily motivated by economic gain \cite{doi:10.1177/00048658211003925,ricaldi_trust}. This includes unauthorized access, misuse, or exchange of digital assets, while politically or ideologically motivated actions are excluded.

The primary infrastructure for these activities is located on the dark web via anonymity networks such as Tor and I2P \cite{10.1145/3655693.3655700}. The ecosystem is structured around two main platform types: \texttt{marketplaces} and \texttt{forums}. Marketplaces enable commoditized exchange of illicit goods using escrow and reputation systems, while forums act as hubs for knowledge exchange, service provision, and network formation \cite{leukfeldt2017cybercriminal}. Although central to anonymous transactions, the ecosystem extends beyond the dark web to the clear web and encrypted messaging platforms such as Telegram for recruitment, coordination, and data leaks \cite{10928256,allodi2024dmitrygoingframingmigratory,ricaldi_trust}. This reflects the distributed and adaptive nature of cybercrime communities.

The World Wide Web is commonly divided into the surface web, deep web, and dark web \cite{Sultana2021}. While the surface web is indexed by search engines and the deep web contains restricted-access content, the dark web requires specialized software to access anonymized services \cite{kaur_dark_2020,Kavallieros2021UnderstandingTheDarkWeb}. From a structural perspective, dark web networks rely on decentralized routing and hub-like connectivity patterns that enable anonymous communication \cite{campobasso2023you,9229679}. While these mechanisms provide privacy protections \cite{Davis2021}, they are also widely used for illicit trade, with marketplaces facilitating transactions through encryption and cryptocurrency-based payments \cite{9738766,Kermitsis2021}.

Prior work shows that dark web content exhibits distinct linguistic and structural characteristics compared to the clear web, including differences in terminology, named entities, and discourse patterns \cite{varghese2023extraction,choshen}. Large-scale analyses further indicate that the dark web forms a structured ecosystem of recurring content categories and domain-specific vocabulary \cite{avarikioti2018structure,jin2022shedding}. These consistent semantic signals have motivated the application of machine learning and natural language processing techniques to analyze dark web content.

\subsection{Analysis Techniques for the Dark Web}\label{ssec:related}
\noindent
The dark web is widely used for illicit activities and has therefore become an important source of CTI \cite{7317717}. Active measurement approaches deploy controlled systems to observe offender interactions and behavior \cite{Ricaldi_AaaS,wang2020into}, yet most research focuses on extracting actionable insights from forums and marketplaces through computational content analysis, i.e., deriving conclusions from textual or multimedia data \cite{klaus_krippendorff_content_2019}. Within cybersecurity, this approach is commonly applied to mine CTI from unstructured sources\cite{de2025methodologies}.

A lot of work treats dark web content as an intelligence source, focusing on extracting structured information such as entities, attack indicators, or threat reports \cite{heistracher2020information,kadoguchi2019exploring,varghese2023extraction}. Reviews further highlight automated extraction of indicators of compromise and behavior patterns \cite{rahman2020literature}. They primarily treat platforms as intelligence feeds rather than evolving socio-technical environments.

Another research direction applies \texttt{NLP} and machine learning to detect or predict cyber threats. However, despite the growing use of AI-driven approaches in CTI, their effective adoption in practice remains limited. 
Recent empirical evidence highlights that challenges such as lack of integration into security workflows, limited trust by analysts, and insufficient robustness and monitoring mechanisms hinder trustworthy deployment \cite{karaosman2026security}. Prior work includes identifying emerging threats in marketplace listings \cite{dong2018new}, predicting exploit occurrences from forum discussions \cite{tavabi2018darkembed}, inferring attacker intent using deep learning \cite{sangher2024lstm,sangher2023towards}, and detecting cyberattacks using sentiment signals \cite{mardassa2024sentiment}.

\texttt{Topic modeling} is another widely used technique, identifying latent semantic themes within large corpora \cite{Jiang2023}. Neural approaches such as BERTopic leverage contextual embeddings and clustering to produce interpretable topics \cite{grootendorst2022bertopicneuraltopicmodeling}, and have been applied to uncover discussion themes in dark web data.

Finally, prior work also focuses on categorization tasks, including classifying marketplace products and forum posts \cite{heistracher2020machine,murty2022dark}, categorizing onion services \cite{ghosh2017automated,pastor2024big,avarikioti2018structure}, and comparing models for detecting illicit activity \cite{dalvi2022comparative,shin2023dark}. Toolkits have also been proposed to identify cybercriminal communities \cite{chen2021amoc}, while sentiment and pattern analysis methods are used to characterize content \cite{murty2021sentiment}. 

\subsection{Research Gap}\label{ssec:research_gap}
\noindent
Prior research has applied NLP and topic modeling techniques to dark web data, but largely outside the scope of economically motivated cybercrime as defined in this study. Early work focuses on uncovering latent themes or communities without centering on cybercriminal activity~\cite{yang2009discovering,LHuillier2011TopicDarkWeb}, while more recent approaches use topic modeling for classification or ontology construction in general threat detection~\cite{shin2023dark,Basheer2024DarkOnto}. Other studies apply these techniques in the context of terrorism and extremism rather than market-oriented cybercrime~\cite{Sonmez2024TerrorismTopic}, and systematic reviews confirm that most work emphasizes broadly defined security threats~\cite{9197590}. Thus, the application of topic modeling to economically motivated cybercrime communities remains limited.

In addition, \textit{studies predominantly analyze dark web data as static snapshots or short-term corpora, providing limited insight into how topics evolve over time across platforms, and longer periods.} To address this gap, this study adopts a longitudinal perspective using repeated HTML snapshots of dark web websites. Specifically, we investigate the following research questions:

\begin{itemize}[leftmargin=*,noitemsep,topsep=0pt]
\item \textbf{RQ1 Topic Prevalence.}  
Which topics are discussed on dark web forums and marketplaces, and how are they distributed across communities?

\item \textbf{RQ2 Topic Lifecycle.}
How do topics change over time in terms of their relative prevalence, persistence, and duration across the observation period?
\end{itemize}
% How do discussion topics evolve over time in terms of their emergence, persistence, and decline?

We conduct a longitudinal analysis of HTML snapshots collected over six years, covering 25,065 dark web websites and 11,403,638 snapshots.

\noindent

% \begin{cooltextbox}
% \textsc{\textbf{Takeaways.}} Use this to highlight cool stuff\footnote{And you can also put footnotes here!}
% \end{cooltextbox}
%\input{sections/0-background}
\section{Methodology}
\label{sec:method}
\noindent
We adopt a snapshot-based longitudinal design to analyze how discussions on dark web platforms evolve over time. The methodology consists of three main stages: (1) preprocessing and structuring HTML snapshots into longitudinal website histories, (2) discovering discussion topics using an embedding-based clustering pipeline, and (3) analyzing topic prevalence and lifecycle over time. An overview of the complete workflow is shown in Figure~\ref{fig:meth} and details on reproducibility are in Appendix \ref{app:reproducibility}.

\begin{figure*}[t]
    \centering
    \includegraphics[width=0.95\linewidth]{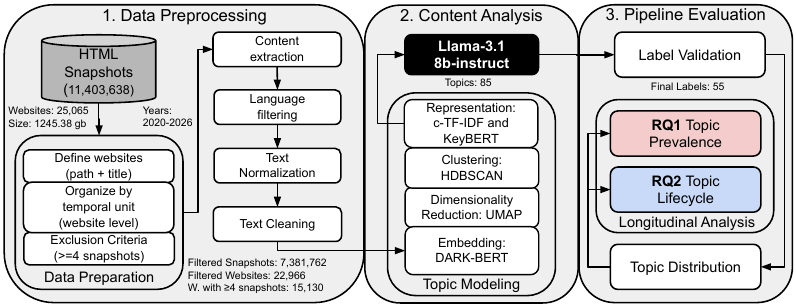}
    \caption{Overview of methodology from data preprocessing, to analysis pipeline, and evaluation}
    \label{fig:meth}
\end{figure*}

\subsection{Dataset and Study Design}
\noindent
The primary unit of our longitudinal analysis is a \textit{website}, defined as a stable webpage identity observed across multiple HTML snapshots. The data used in this study was collected with the Dark Web Monitor Tool by CFLW Cyber Strategies, an organization specializing in cybersecurity and dark web monitoring. Their services support law enforcement agencies and public institutions\footnote{All data utilized in this study can be obtained by researchers after submitting a research proposal with IRB approval.}.

The dataset consists of \textbf{11,337,512 HTML snapshots} (approximately \textbf{1245.38 GB}) collected over six years (\textbf{2020–2026}\footnote{Earliest snapshot timestamp: 2020-01-01 14:37:52; Latest snapshot timestamp: 2026-01-30 19:13:34.}). Each snapshot represents a captured HTML version of a ``dark'' web webpage at a specific point in time. This longitudinal archive allows us to observe how content on dark web platforms changes over repeated observations of the same website. More details on the dataset are provided in Appendix \ref{app:filtering and temporal}.

\subsection{Data Preprocessing} \label{ssec:preprocessing}

\noindent Before analysis, raw HTML snapshots were processed to construct longitudinal website histories. The preprocessing stage consists of website grouping, snapshot filtering, content extraction and text normalization.

\vspace{-3mm}
\subsubsection{Website Grouping and Snapshot Exclusion}

Snapshots were grouped based on a combination of file path and page title, which together provide a robust identifier for the same webpage across time. Each group represents a single website observed at multiple points in time and was assigned a unique internal identifier. Duplicate and near-duplicate snapshots were removed during preprocessing.

Within each group, snapshots were sorted chronologically using their creation timestamps. Groups containing four or less snapshots were removed, as they do not provide sufficient temporal depth to analyze content evolution. Additionally, metadata entries without corresponding HTML files were excluded to ensure dataset consistency.

\vspace{-3mm}
\subsubsection{Content Extraction and Text Normalization}

Raw HTML snapshots were converted into plain text to isolate meaningful webpage content and remove structural noise. HTML parsing and content extraction were performed using the \texttt{trafilatura} library,\footnote{Trafilatura is specifically designed for web content extraction and reliably removes navigation elements, scripts and boilerplate HTML structures.} which extracts the main textual content of web pages while excluding scripts, navigation menus and other non-informational elements. To ensure linguistic consistency, extracted text was filtered by language using the \texttt{langdetect} library. Only English-language snapshots were retained for analysis in order to avoid any noise into the topic modeling process.

The extracted text was then normalized through lowercasing and removal of URLs, email addresses, punctuation, special characters, numbers, and repeated characters. Tokens were lemmatized using the \texttt{WordNet} lemmatizer to reduce words to their base forms. Stopwords were removed and tokens shorter than three characters or longer than 25 characters were discarded\footnote{Such unusually long strings are typically non-linguistic artifacts (e.g., concatenated words, encoding errors or residual URLs) that do not contribute meaningful semantic information and may negatively impact embedding quality and topic coherence.}. The resulting corpus represents a standardized and noise-reduced textual dataset suitable for embedding generation and topic modeling. Details on dataset filtering, inclusion criteria, and temporal handling are documented in Appendix \ref{app:filtering and temporal}.

Individual HTML snapshots may contain multiple discussion threads or mixed topics, particularly in forum index pages or aggregated marketplace listings. In our approach, each snapshot is treated as a single document and represented as a probabilistic mixture over topics. Subthread-level segmentation is not performed, which may lead to blended topic representations in pages containing heterogeneous content. This limitation is further discussed in \ref{ssec:limitations} and represents an avenue for future work. 

% \subsubsection{Corpus Validation}

% To verify the correctness of the longitudinal website construction, a stratified sample of snapshot groups was manually reviewed using a predefined validation protocol. The review focused on confirming that snapshots grouped under the same identifier correspond to the same webpage across time.

% Annotators inspected sampled snapshot groups to verify that they represented consistent webpage identities, such as the same forum thread or marketplace page observed at different time points. Particular attention was given to detecting potential crawler artifacts, including cases where different pages might have been incorrectly grouped due to similar titles or file paths.

% This validation step ensures that the grouping procedure reliably reconstructs longitudinal website histories and that observed temporal changes reflect genuine updates of the same page rather than accidental mixing of unrelated webpages. Details of the validation procedure are provided in Appendix~\ref{app:human-validation}.

\subsection{Content Analysis Pipeline} \label{ssec:topic_discovery}

\noindent After preprocessing, each snapshot is passed through a structured topic discovery pipeline that converts textual content into interpretable thematic categories. The pipeline follows a sequential representation–clustering–labeling workflow illustrated in Figure~\ref{fig:meth}.

\vspace{-3mm}
\subsubsection{Semantic Embedding}

Each cleaned snapshot is converted into a semantic vector representation using contextual embeddings. We employ DARK-BERT \cite{jin2023darkbert}, a domain-adapted transformer model trained on cybersecurity and darknet-related corpora, to generate document embeddings. Domain-specific embedding models have been shown to better capture specialized vocabulary and semantic relationships in technical security discussions.

Other embedding models, including general and cybersecurity models, were evaluated during pipeline development. DARK-BERT consistently produced more coherent topic structures and semantically consistent document clusters and after manual verification was selected for the final pipeline. The model selection procedure and comparative evaluation are described in Appendix~\ref{app:parameter-evaluation}.

\vspace{-3mm}
\subsubsection{Dimensionality Reduction}

Because transformer embeddings are high-dimensional, we use Uniform Manifold Approximation and Projection (UMAP) to project embeddings to lower-dimensional space while preserving semantic neighborhood structure. This improves clustering stability and separates semantically distinct discussions while keeping relationships between related activities.

\vspace{-3mm}
\subsubsection{Topic Clustering}

Topics are discovered using density-based clustering (HDBSCAN). Unlike centroid-based clustering methods, HDBSCAN does not require specifying the number of clusters in advance and can identify dense thematic clusters while treating sparse content as noise. Each resulting cluster corresponds to a candidate discussion topic representing a recurring theme.

\vspace{-3mm}
\subsubsection{Topic Representation and Labeling}

To interpret the discovered clusters, representative keywords are extracted using class-based TF-IDF, which identifies terms that distinguish each cluster from the rest of the corpus. KeyBERT is additionally used to extract representative keyphrases that improve semantic readability.

Because keyword lists can remain ambiguous for large-scale analysis, clusters are assigned semantic topic labels using the large language model (LLM) \textit{Llama-3.1-8B-Instruct}. While LLMs enable scalable and interpretable labeling of large text corpora, their use in security-critical contexts must be carefully considered, as they can introduce reduce independent reasoning in human decision-making \cite{pekaric2025llms}. Thus, the model receives representative keywords and example documents from a cluster to generate a clear and concise human-readable topic name.

For example, one cluster produced the keywords \textit{product}, \textit{bought}, \textit{vendor}, \textit{shipping}, \textit{escrow}, and \textit{bitcoin}. These terms reflect transactional discussions commonly found on illicit marketplaces. Based on these keywords and example documents, the language model generated the label \textit{Online Shopping}. This process converts machine-generated keyword sets into interpretable thematic labels suitable for longitudinal analysis.

All generated labels were manually validated to ensure semantic accuracy and consistency. Labels that were similar, or a subset of another (e.g., label "Market Transactions" aggregated to "Online Shopping") were merged to ensure proper representation. Details of the validation and label merging protocol are provided in Appendix~\ref{app:labelling_topics}.

\subsection{Longitudinal Topic Analysis} \label{ssec:longitudinal}

To analyze thematic time dynamics, each snapshot is represented as a probability distribution over discovered topics using BERTopic's soft assignment mechanism. Rather than assigning a single topic to each snapshot, this approach captures mixtures of topics within a document.

These topic distributions are then aggregated across snapshots belonging to the same website and time interval, producing time-indexed topic prevalence measures. This representation allows gradual thematic changes to be measured over time without forcing abrupt topic switches. Two quantitative measures are derived from these temporal topic distributions to address the research questions.

\noindent \textbf{Topic Prevalence (RQ1).}  
\noindent Topic prevalence is measured as the aggregated topic probability mass across snapshots within each time interval. This metric quantifies how strongly each topic is represented across the broader ecosystem of cybercrime forums and marketplaces.

\noindent \textbf{Topic Lifecycle (RQ2).}  
Topic lifecycle is measured using temporal change indicators, including topic lifespan, growth and decay rate. These metrics capture how topics emerge, persist and decline over the observed period.

\subsection{Validation and Reliability} \label{ssec:validation}

\noindent To ensure the semantic validity and consistency of the discovered topics, a manual validation process was conducted. Two independent annotators reviewed the automatically generated topic labels based on representative keywords and documents. Annotators assessed whether each label accurately reflected the underlying content and reworded or refined labels where necessary. In addition, they independently evaluated potential overlap between topics and merged them when substantial semantic similarity was identified. We provide further details on the validation procedure including examples in Appendix~\ref{app:human-validation}.

\subsection{Ethical Considerations}

\noindent This study analyzes textual content from publicly accessible dark web forums and marketplaces. No interaction with users, accounts, or services occurred, and no attempts were made to access restricted areas, in line with established internet-mediated research guidelines \cite{hewson2013ethics}. Given the potential presence of illicit or sensitive material, only text necessary for aggregate analysis was processed, and no individuals or accounts were profiled. Results are reported exclusively at the aggregated level, focusing on ecosystem-wide patterns \cite{pickering2021ethical}. All analysis was conducted offline on archived snapshots, and no sensitive operational details (e.g., access methods, credentials, or live URLs) are disclosed. The dataset was preprocessed by the data provider to remove personally identifiable information. No direct threats or actionable intelligence targeting specific individuals were identified, and no reporting to authorities was required.

\section{Results}

\noindent We present the results of the longitudinal topic analysis. 
The topic discovery pipeline outputted
\textbf{85} distinct discussion topics extracted from 
\textbf{7,381,762}a HTML snapshot. For each label produced by Llama (e.g., Databases, Forum Security, Prepaid Cards), topics were inspected to assess whether they reflected a consistent and interpretable theme and were distinguishable from other topics. After label verification and correction, they were aggregated by merging similar labels, which resulted in \textbf{55} final labeled topics.

To improve interpretability, we also group the labels into four overarching categories based on their semantic structure and functional roles within the ecosystem (see Table \ref{tab:aggregated_topics}). The \textit{Transactional} category captures activities related to financial exchange, including payment mechanisms, transaction security, and monetization processes. The \textit{Products} category represents the supply side of the ecosystem, covering the trade of illicit goods and services. The \textit{Infrastructure} category includes the technical and operational components that enable cybercriminal activity, such as tools and supporting technologies. Finally, the \textit{Community} category captures the social and informational layer of the ecosystem, including coordination, trust-building and knowledge sharing among participants, primarily within forum environments.

%across \textbf{22,966} dark web websites

%Across the dataset, the resulting topics capture a broad range of  economically motivated cybercrime activities, including marketplace transactions, service offerings, infrastructure discussions, and community interactions.

%The results indicate that topics within the same cluster share domain-specific terminology and contextual meaning, supporting the effectiveness of the embedding, clustering, and labeling pipeline.

The remainder of this section addresses the research questions. 
First, we analyze how topics are distributed across platforms and how prevalent they are (§\ref{ssec:results_rq1}), and then examine their emergence and decay through time (§\ref{ssec:results_rq2}).

%\subsection{Thematic Aggregation of Labels}

%\noindent To improve interpretability, we group the labels into four overarching categories based on their semantic structure and functional roles within the ecosystem (see Table \ref{tab:aggregated_topics}). The \textit{Transactional} category captures activities related to financial exchange, including payment mechanisms, transaction security, and monetization processes. The \textit{Products} category represents the supply side of the ecosystem, covering the trade of illicit goods and services. The \textit{Infrastructure} category includes the technical and operational components that enable cybercriminal activity, such as tools and supporting technologies. Finally, the \textit{Community} category captures the social and informational layer of the ecosystem, including coordination, trust-building and knowledge sharing among participants, primarily within forum environments.

\subsection{Topic Prevalence (RQ1)} \label{ssec:results_rq1}

\noindent First, we analyze how discussion topics are distributed across dark web marketplaces and forums, and how their prominence varies across the observed period. 

The aggregated distribution shows that community (47.38\%) and transactional (20.59\%) activities dominate the ecosystem, together accounting for the majority of observed discussion volume, while infrastructure (13.65\%) and product-related topics (6.75\%) represent smaller portions of the corpus, as shown in Table \ref{tab:aggregated_topics}.

\begin{table}[t]
\centering
\caption{Top 20 topics labels by corpus share, mapped to category}
\label{tab:aggregated_topics}
\begin{tabular}{p{0.8cm} p{3.5cm} p{1.1cm} p{0.9cm}}
\toprule
\textbf{Category} & \textbf{Topic Labels} & \textbf{S. Count} & \textbf{\%Corp.} \\
\midrule

\multirow{6}{*}{\rotatebox[origin=c]{90}{Transactional}} 
& Online Shopping & 558,658 & 8.54 \\
& Transaction Protection & 343,545 & 5.26 \\
& Online Banking & 175,188 & 2.68 \\
& Money Making Opportunities & 142,684 & 2.18 \\
& Prepaid Cards & 85,679 & 1.31 \\
& Vendor Sales \& Shipping & 90,145 & 1.38 \\
\midrule

\multirow{3}{*}{\rotatebox[origin=c]{90}{Prod.}} 
& Stolen Bank and Payment Acc. & 147,206 & 2.25 \\
& Forged Document Services & 146,661 & 2.24 \\
& Counterfeit Money & 147,332 & 2.25 \\
\midrule

\multirow{4}{*}{\rotatebox[origin=c]{90}{Infras.}} 
& Infrastructure and Hosting & 308,950 & 4.73 \\
& VPN Services & 138,165 & 2.11 \\
& Databases & 190,900 & 2.92 \\
& Browser Hijacking & 92,592 & 1.42 \\
\midrule

\multirow{7}{*}{\rotatebox[origin=c]{90}{Community}} 
& Torrents and Files & 1,027,774 & 15.71 \\
& Forum Reputation & 904,299 & 13.83 \\
& Forum Features & 524,838 & 8.02 \\
& Forum Security & 117,604 & 1.80 \\
& Vendor Channels and Prom. & 76,691 & 1.17 \\
& WikiLeaks & 209,978 & 3.21 \\
& Politics and War & 235,484 & 3.60 \\
\bottomrule
\end{tabular}
\end{table}

The most prevalent topics are \textit{Torrents and Files} (15.71\%) and \textit{Forum Reputation} (13.83\%), followed by \textit{Online Shopping} (8.54\%) and \textit{Forum Features} (8.02\%). Beyond these leading topics, the remaining distribution consists of a larger number of lower-frequency topics, each contributing less than 8\% of the corpus. The least prevalent topics in the top 20 set include \textit{Browser Hijacking} (1.42\%) and \textit{Vendor Sales \& Shipping} (1.38\%), which are still considerably higher when compared to \textit{Forum Credentials} (0.01\%), \textit{Digital Payment Card Services} (0.03\%), \textit{Leaked Corporate Files} (0.03\%), and \textit{Banned Accounts} (0.03\%) which were the least prevalent out of the 55 labels. The complete set of 55 topic labels is shown and defined in Appendix \ref{app:topics}.

%Overall, the distribution is uneven, with a small number of topics accounting for a substantial share of the corpus and a long tail of less frequent topics.

\vspace{-3mm}
\subsubsection{Distribution Across Platforms}

\begin{figure}[t]
    \centering
    \includegraphics[width=\columnwidth]{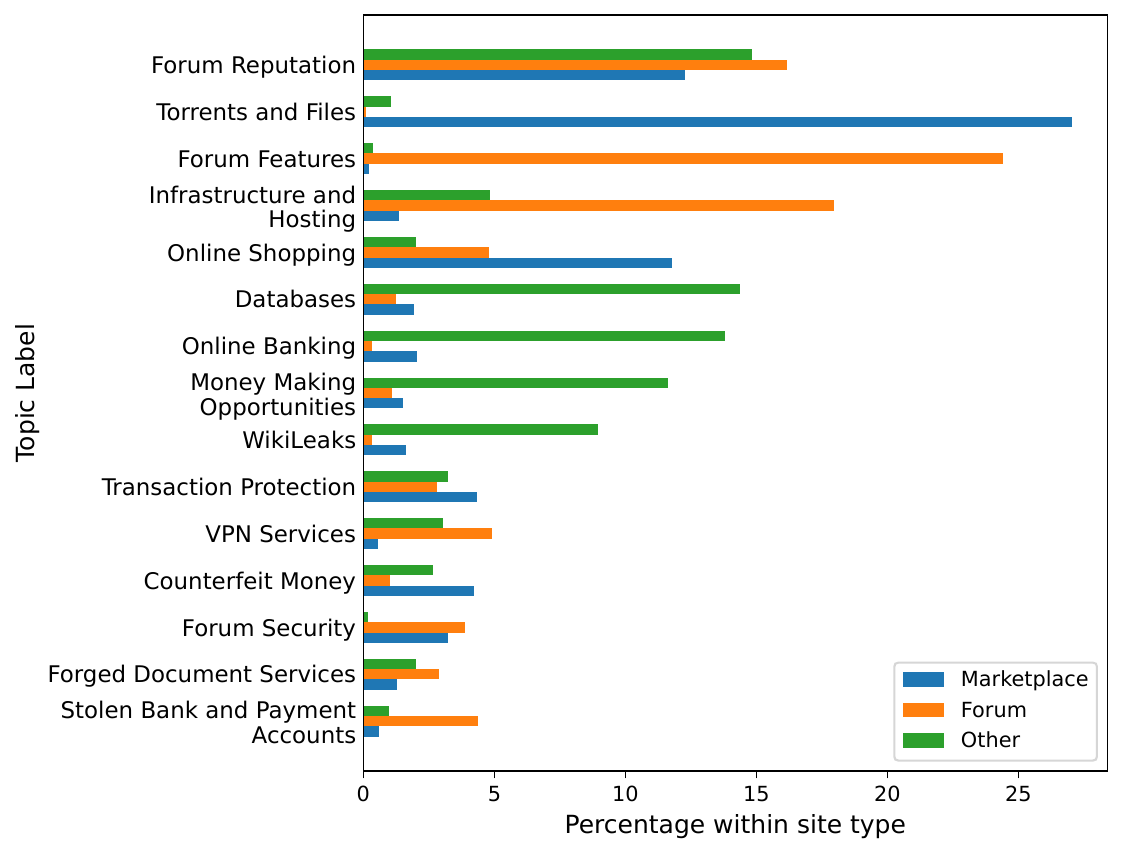}
    \caption{Distribution of marketplace vs forum vs other of top 20 topic labels in corpus.}
    \label{fig:marketplace_forum_other}
\end{figure}

Topic prevalence differed substantially between forums and marketplaces. \autoref{fig:marketplace_forum_other} compares topic composition across marketplaces, forums, and other site types. Naturally, forum-related topics such as \textit{Forum Reputation}, \textit{Forum Features}, and \textit{Forum Security} are more prominent in forums, while transaction- and product-related topics such as \textit{Online Shopping}, \textit{Stolen Bank and Payment Accounts}, and \textit{Forged Document Services} show higher representation in marketplaces.

Several topics, including \textit{Infrastructure and Hosting} and \textit{Online Shopping}, appear across multiple site types with varying proportions. Only few topics represent a somewhat even share across platforms, amongst them, \textit{Forum Reputation} and \textit{Transaction Protection}. Overall, the distribution indicates differences in topic composition between platform types, with certain topics concentrated more strongly in specific environments.

% FIGURE TO ADD:
% Horizontal bar chart
% X-axis: Percentage of corpus
% Y-axis: Topic category
% Shows overall ecosystem composition

% In general, the ecosystem was dominated by a relatively small set of highly 
% prevalent topics, while a long tail of less frequent topics represented 
% a smaller share of discussion activity.

\vspace{-3mm}
\subsubsection{Temporal Prevalence}

%Topic prevalence varied over time. Rather than abrupt thematic shifts, the ecosystem exhibited gradual changes in the prominence of existing topics across the observation period. \autoref{fig:temporal_prelevance} presents the temporal prevalence of the top 20 topics across the observation period. Several topics, including \textit{Forum Reputation} and \textit{Online Shopping}, maintain a visible presence throughout multiple time intervals.

\begin{figure}[t]
    \centering
    \includegraphics[width=\columnwidth]{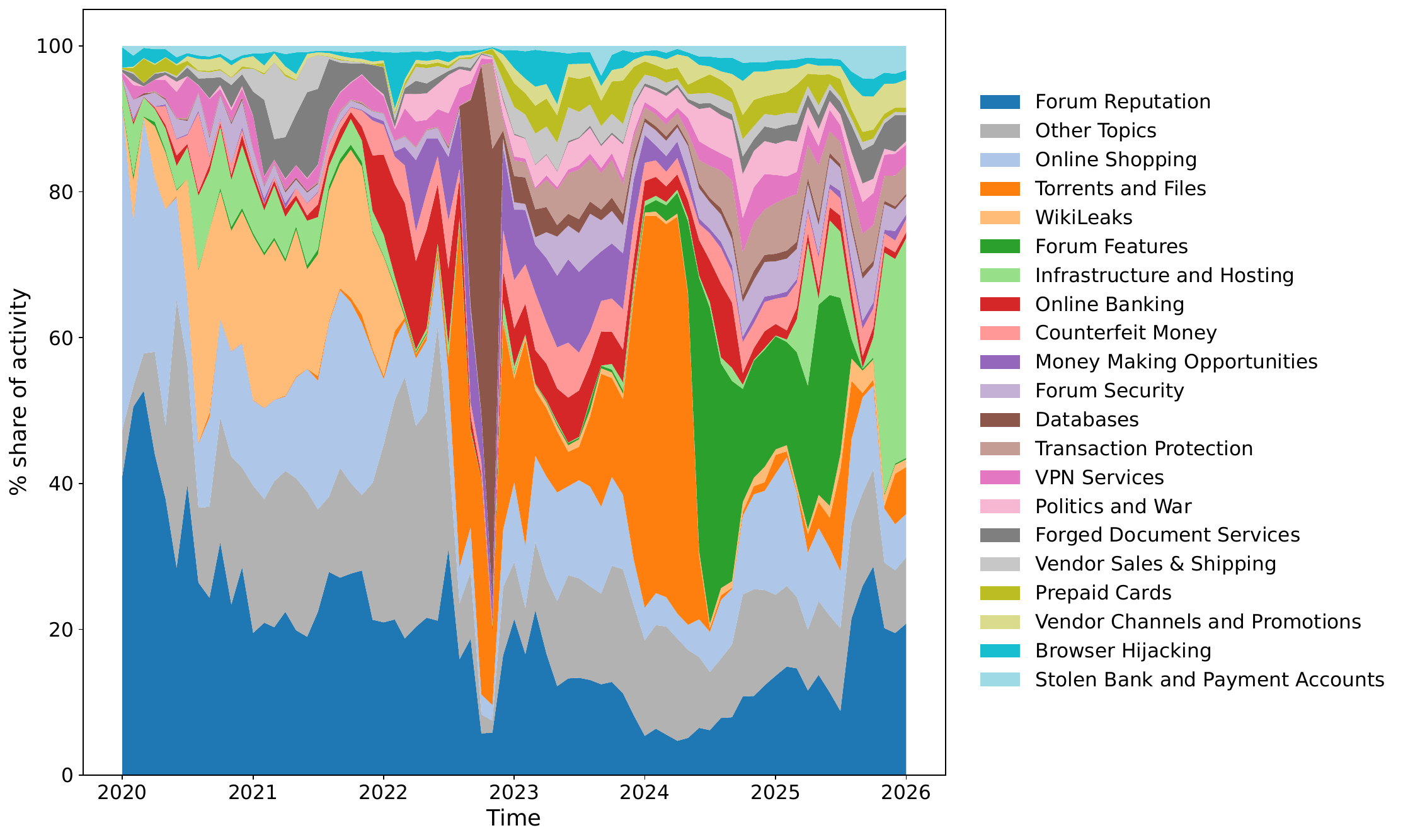}
    \caption{Temporal Prevalence of Top 20 Topics.}
    \label{fig:temporal_prelevance}
\end{figure}

Topic prevalence evolves over time, wherein the changes are gradual rather than disruptive. As shown in Figure \ref{fig:temporal_prelevance}, the top 20 topics persist across the observation period, with shifts primarily occurring in their relative prominence (not through new theme introduction). Core topics such as \textit{Forum Reputation} and \textit{Online Shopping} remain consistently visible across multiple intervals.

Overall, the ecosystem is characterized by stability within a fixed thematic structure. While individual topics rise or decline in importance, the dominant set of topics remains largely unchanged. This suggests that temporal dynamics are driven by a redistribution of attention within an established core and not by continuous thematic turnover. Within this stable structure, we observe notable variations in prominence. Between 2020 and 2022, \textit{Forum Reputation} dominated the discourse, followed by \textit{WikiLeaks}, \textit{Online Shopping}, and \textit{Infrastructure and Hosting}, with \textit{Online Banking} gaining relative importance toward the end of the period.

%This aligns with our previously observed concentration effects, where a small number of topics account for the majority of discussion.

\vspace{-3mm}
\subsubsection{Quantitative prevalence indicators} The top five topics account for 53.15\% of total discussion volume, the top ten account for 68.67\%, and the top twenty account for 88.36\% of the corpus. Only five topics are required to explain 50\% of all observed activity, and thirteen topics explain 75\%. This shows that topic prevalence is not broadly distributed across many equally important themes, but ordered around a small set of dominant activities.

The concentration pattern is reinforced by recurrence behavior. At the level of the final 55 grouped topics, all identified themes are either continuous or recurring, meaning that no final topic cluster appears only once across the observation window. Thus, the dominant topics are not only highly prevalent, but also structurally persistent over time. This indicates that the dark web ecosystem is organized around stable thematic functions. 

%rather than a large number of short-lived dominant themes.

\begin{cooltextbox}
    \textsc{\textbf{Answer to RQ1:}}
    Topic prevalence is highly concentrated in a small number of dominant themes. Five topics explain more than half of total discussion volume, and thirteen topics explain three quarters of the corpus. At the same time, these dominant topics remain persistent across the observation window, which indicates that the ecosystem is structured around a stable set of recurring functions rather than a broad or rapidly changing distribution of themes.
\end{cooltextbox}

%At the same time, the relative share of individual topics varies over time, with some topics increasing or decreasing in prominence across specific periods. From 2020 to 2022, \textit{Forum Reputation} drove the discourse, followed by \textit{WikiLeaks}, \textit{Online Shopping} and \textit{Infrastructure and Hosting}, with \textit{Online Banking} gaining some prevalence by the end. 

%From 2022 to 2024, \textit{Torrents and Files}, \textit{Databases}, and \textit{Money Making Opportunities} start taking a more prominent share in the corpus. Various topics gain prevalence between 2023 to 2024 particularly, makign the share of the corpus less monopolized. No single topic dominates the entire timeline, and multiple topics contribute to the overall activity in each interval. From 2024 to 2025, \textit{Forum Features} emerges and becomes quite prevalent, signaling a phenomenon worth investigating.

% Some themes maintained stable levels of activity throughout multiple years, 
% while others increased or decreased in relative prominence during specific 
% periods. 
% These patterns indicate that dark web communities tend to evolve through incremental changes in discussion focus rather than sudden reorganization.

% FIGURE TO ADD:
% Stacked area chart
% X-axis: time (months)
% Y-axis: % share of activity
% Color: topic category

\subsection{Topic Lifecycle (RQ2)} \label{ssec:results_rq2}

\noindent While prevalence captures how strongly topics are represented within the ecosystem, lifecycle analysis examines how long topics remain active and how their activity levels change over time.

\vspace{-3mm}
\subsubsection{Topic Lifespan Distribution}

\begin{figure}[t]
    \centering
    \includegraphics[width=\columnwidth]{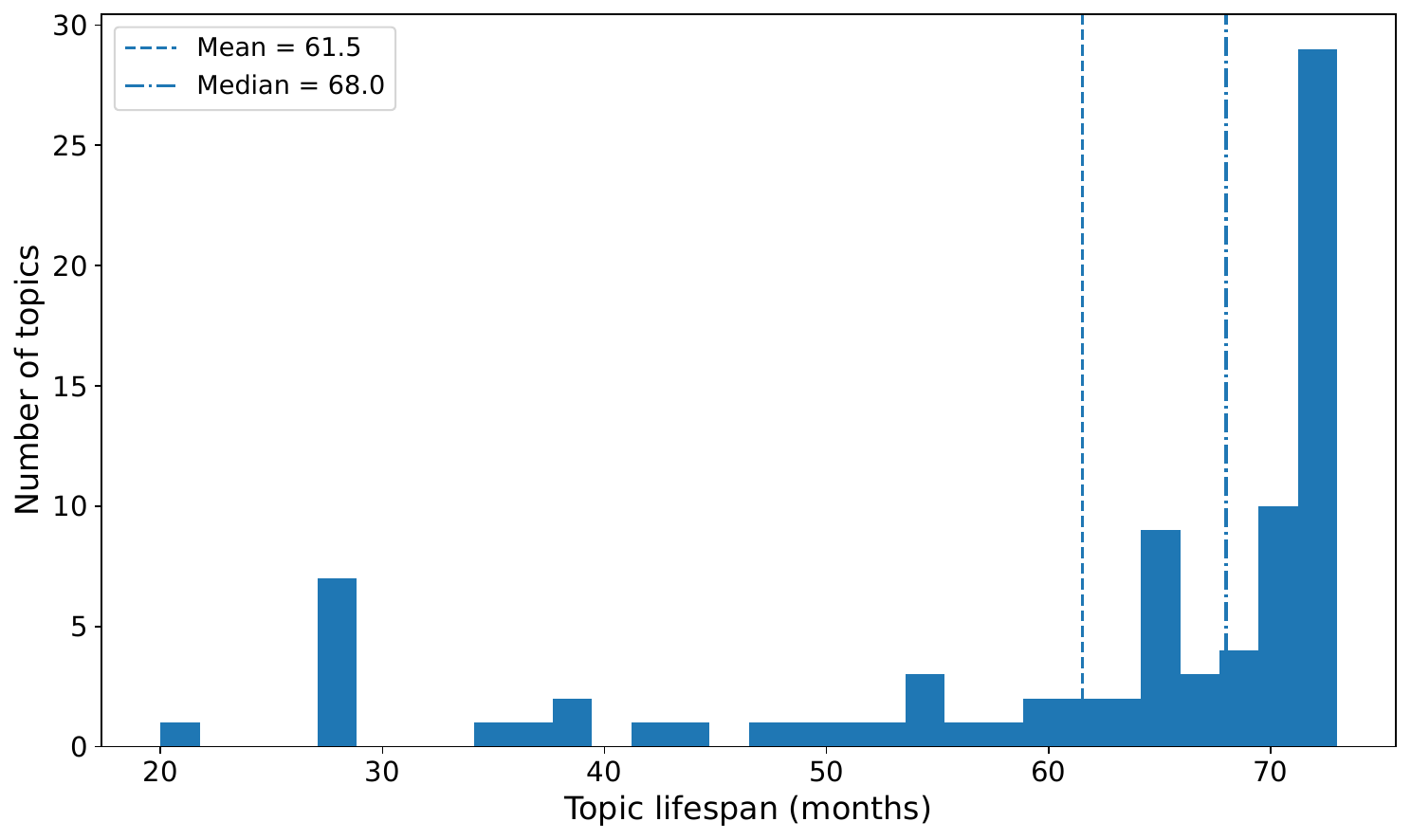}
    \caption{Topic Lifespan Distribution.}
    \label{fig:rq2_topic_distribution}
\end{figure}

Figure~\ref{fig:rq2_topic_distribution} shows the distribution of topic lifespans measured in months across the observation period. The distribution is strongly skewed toward longer durations, with most topics remaining active for a substantial portion of the six-year timeframe. The median lifespan is approximately 68 months, while the mean lifespan is 61.5 months, indicating that the majority of topics persist over multiple years.

Only a small number of topics exhibit shorter lifespans, suggesting that short-lived discussions are relatively uncommon. Instead, the ecosystem is dominated by long-standing topics that remain continuously present, with variation occurring primarily in their level of activity. This pattern indicates that topic turnover is limited and that most thematic structures persist over time.

\vspace{-3mm}
\subsubsection{Temporal Activity and Persistence}

\begin{figure}[t]
    \centering
    \includegraphics[width=\columnwidth]{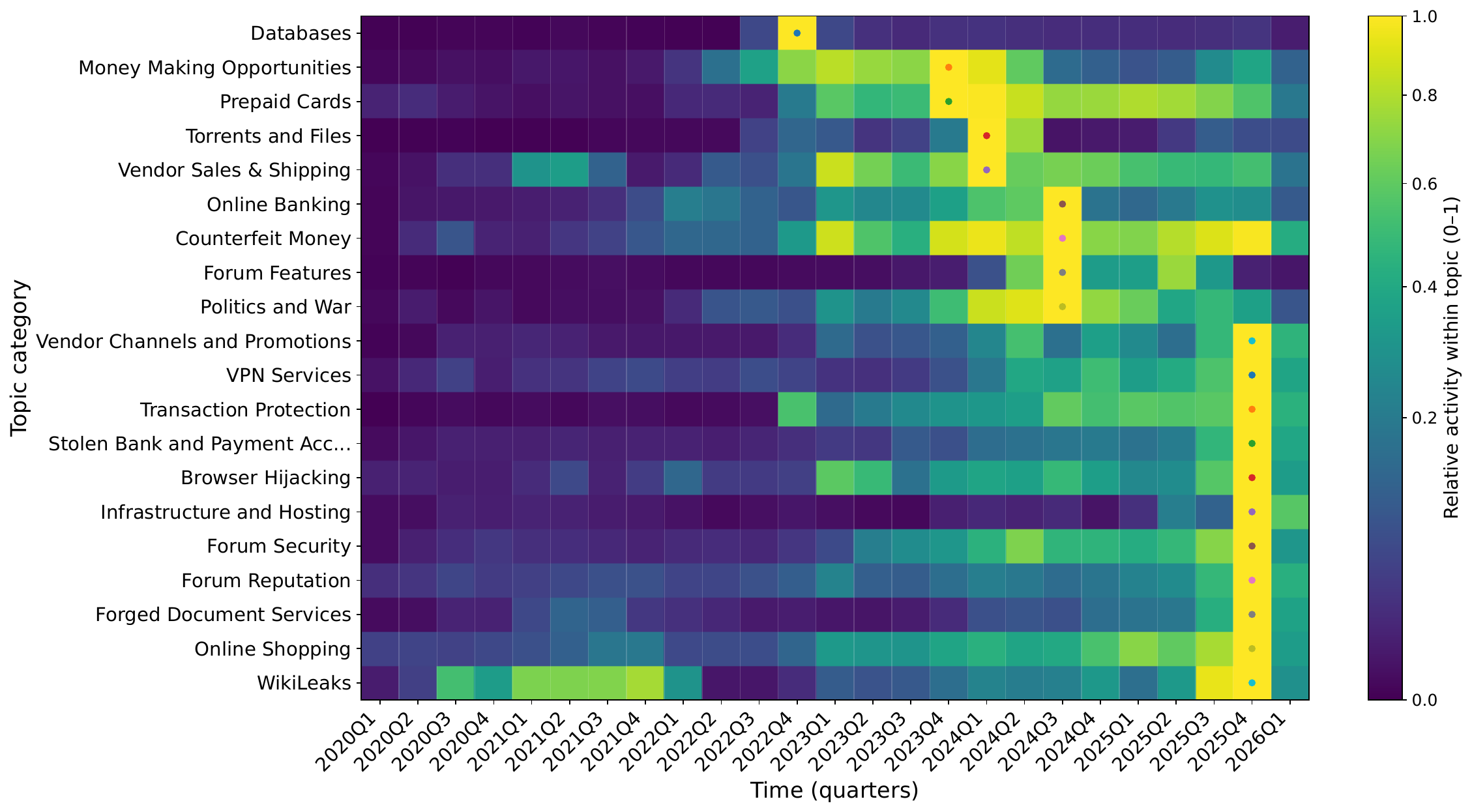}
    \caption{Topic Lifecycle.}
    \label{fig:rq2_topic_lifecycle}
\end{figure}

\begin{figure}[t]
    \centering
    \includegraphics[width=\columnwidth]{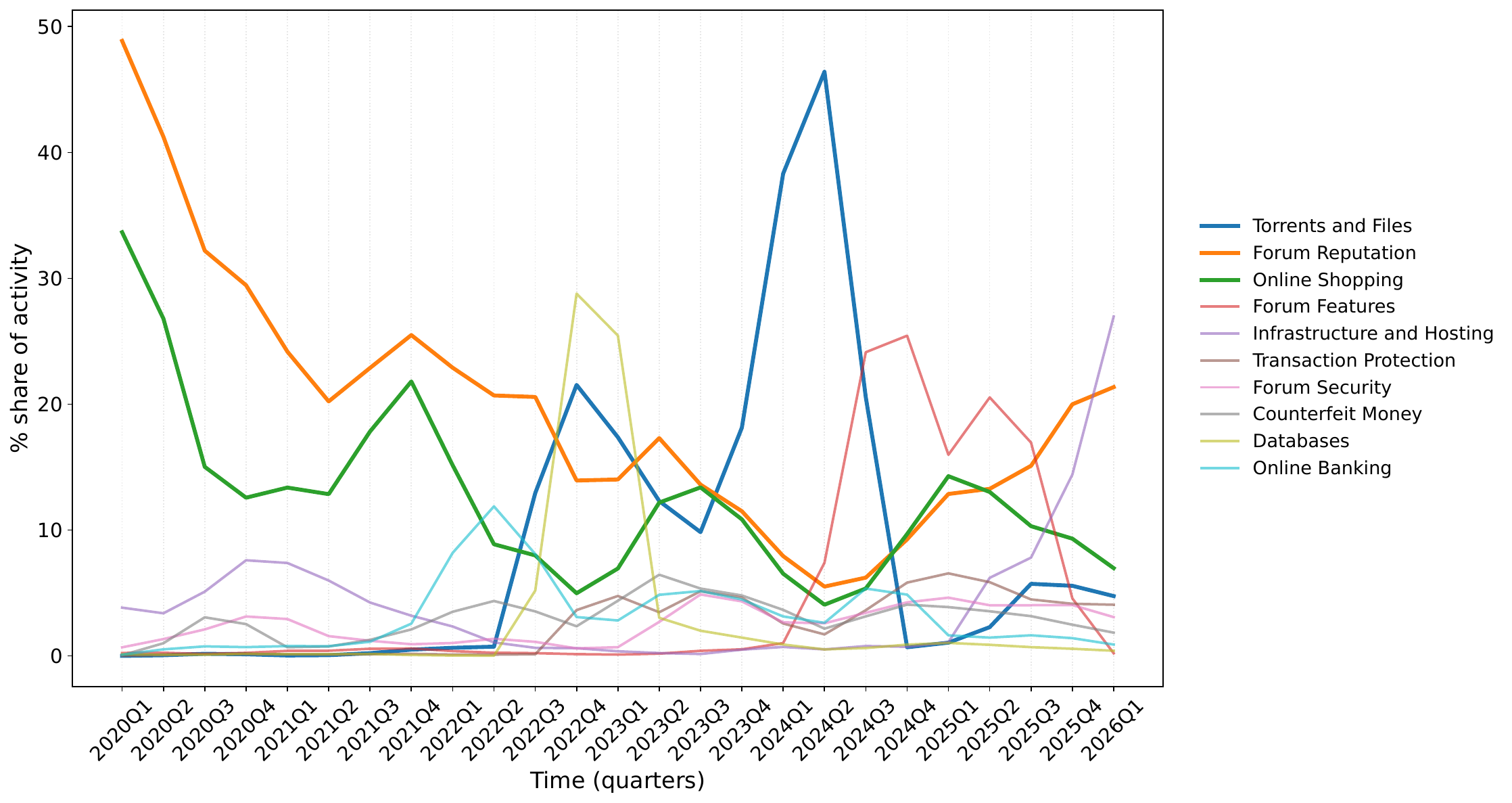}
    \caption{Topic Temporal Prevalence (Top ten Topics).}
    \label{fig:topic_temporal_prevalence_top_ten}
\end{figure}

Most topics remain active across large portions of the observation period, indicating persistent thematic structures. Variation is visible primarily in relative intensity rather than in complete disappearance. Figure~\ref{fig:rq2_topic_lifecycle} shows that activity is redistributed across topics over time, but the underlying topic set remains largely stable. Figure~\ref{fig:topic_temporal_prevalence_top_ten} further illustrates that even the most prevalent topics follow smooth trajectories rather than abrupt discontinuities.

The concentration of activity in long-lived topics implies that cybercrime communities are not defined by rapid thematic churn. Instead, the same core activities remain active for longer periods, while their relative prominence changes in response to broader contextual conditions.

\vspace{-3mm}
\subsubsection{Growth and Decay Dynamics}

\begin{figure}[t]
    \centering
    \includegraphics[width=\columnwidth]{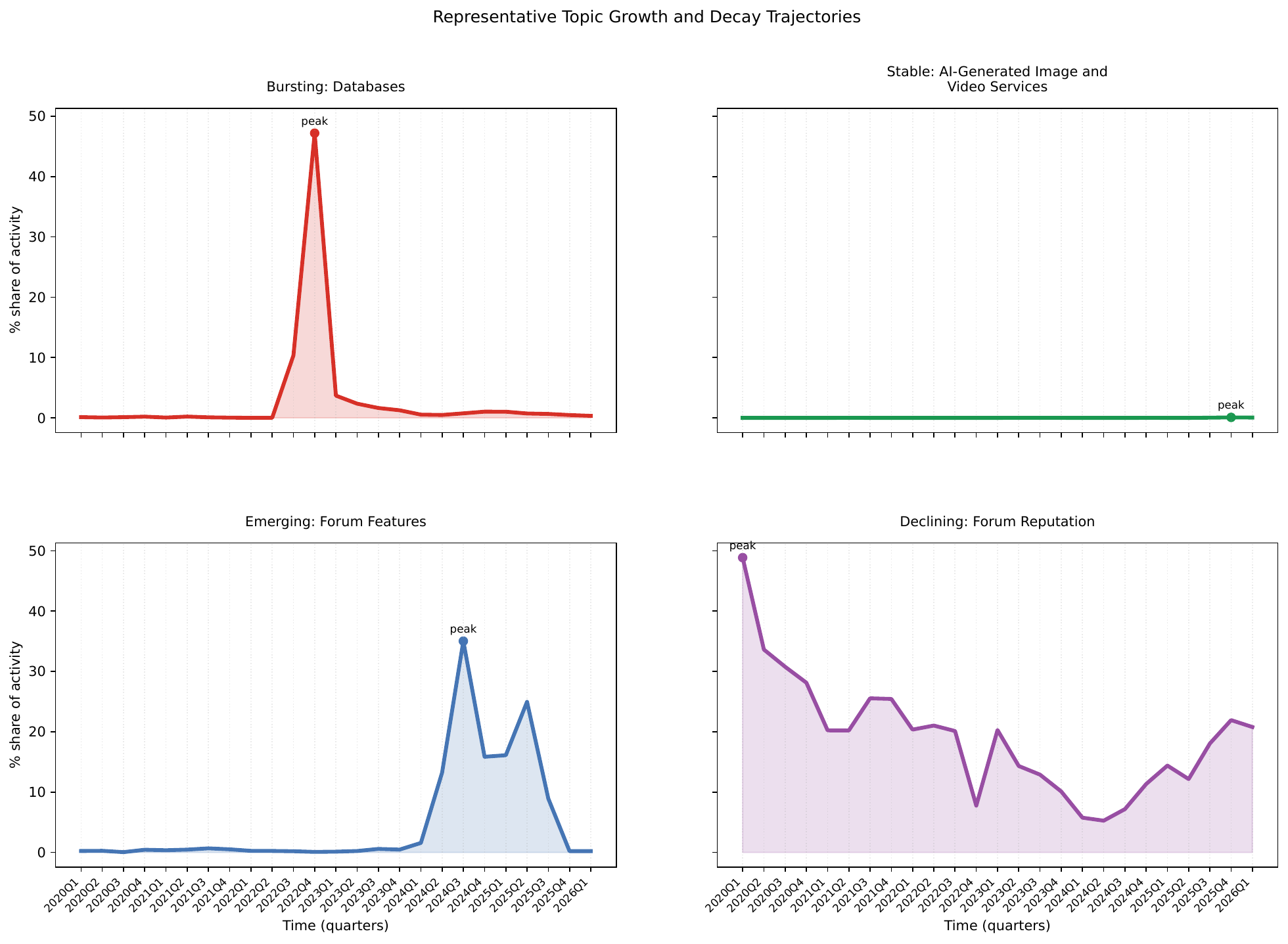}
    \caption{Representative Growth and Decay of Topics.}
    \label{fig:rq2_growth_decay_small}
\end{figure}

\begin{figure}[t]
    \centering
    \includegraphics[width=\columnwidth]{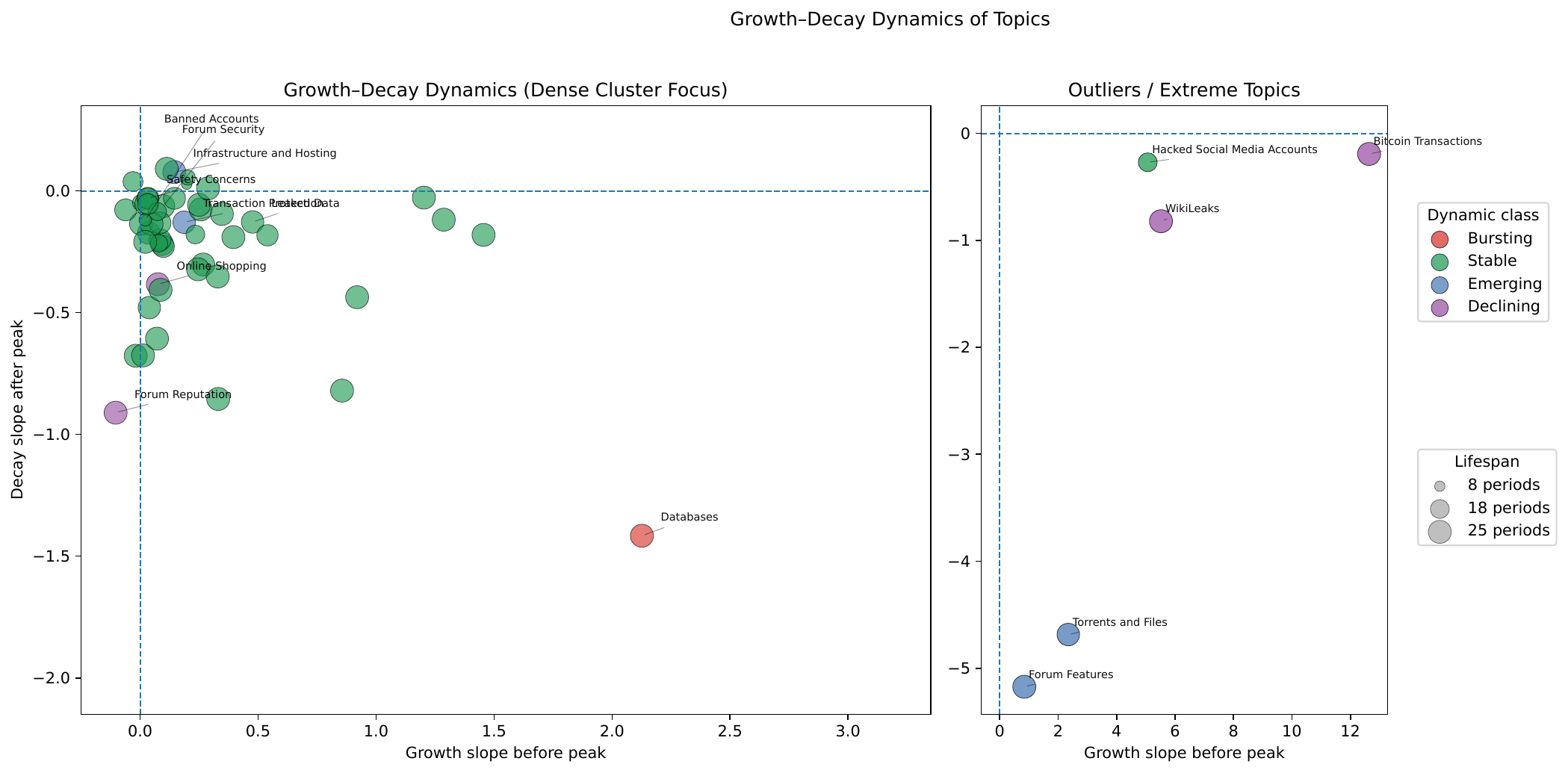}
    \caption{Growth and Decay Dynamics of Topics.}
    \label{fig:rq2_growth_decay_scatter}
\end{figure}

Figure~\ref{fig:rq2_growth_decay_small} shows representative growth and decay trajectories across topics. Distinct patterns are observed. \textit{Databases} exhibits a bursting pattern, with a rapid increase to a clear peak followed by a relatively sharp decline. In contrast, \textit{AI-Generated Image and Video Services} remains stable, maintaining a relatively constant activity level over time. \textit{Forum Features} shows an emerging pattern, with gradual growth across multiple periods leading to a late peak. Conversely, \textit{Forum Reputation} displays a declining pattern, with higher early activity followed by a steady decrease.

Figure~\ref{fig:rq2_growth_decay_scatter} summarizes these dynamics using growth and decay slopes. Most topics cluster in a central region, indicating moderate growth and moderate decay. This includes topics such as \textit{Forum Reputation}, \textit{Online Shopping}, and \textit{Infrastructure and Hosting}, which exhibit balanced and gradual lifecycle dynamics.

A small number of topics appear as outliers. \textit{Torrents and Files} and \textit{Forum Features} show higher growth slopes, indicating faster increases before peak, combined with stronger post-peak declines. In contrast, more stable topics are located near the origin, reflecting limited variation in both growth and decay.

\begin{figure}[t]
    \centering
    \includegraphics[width=\columnwidth]{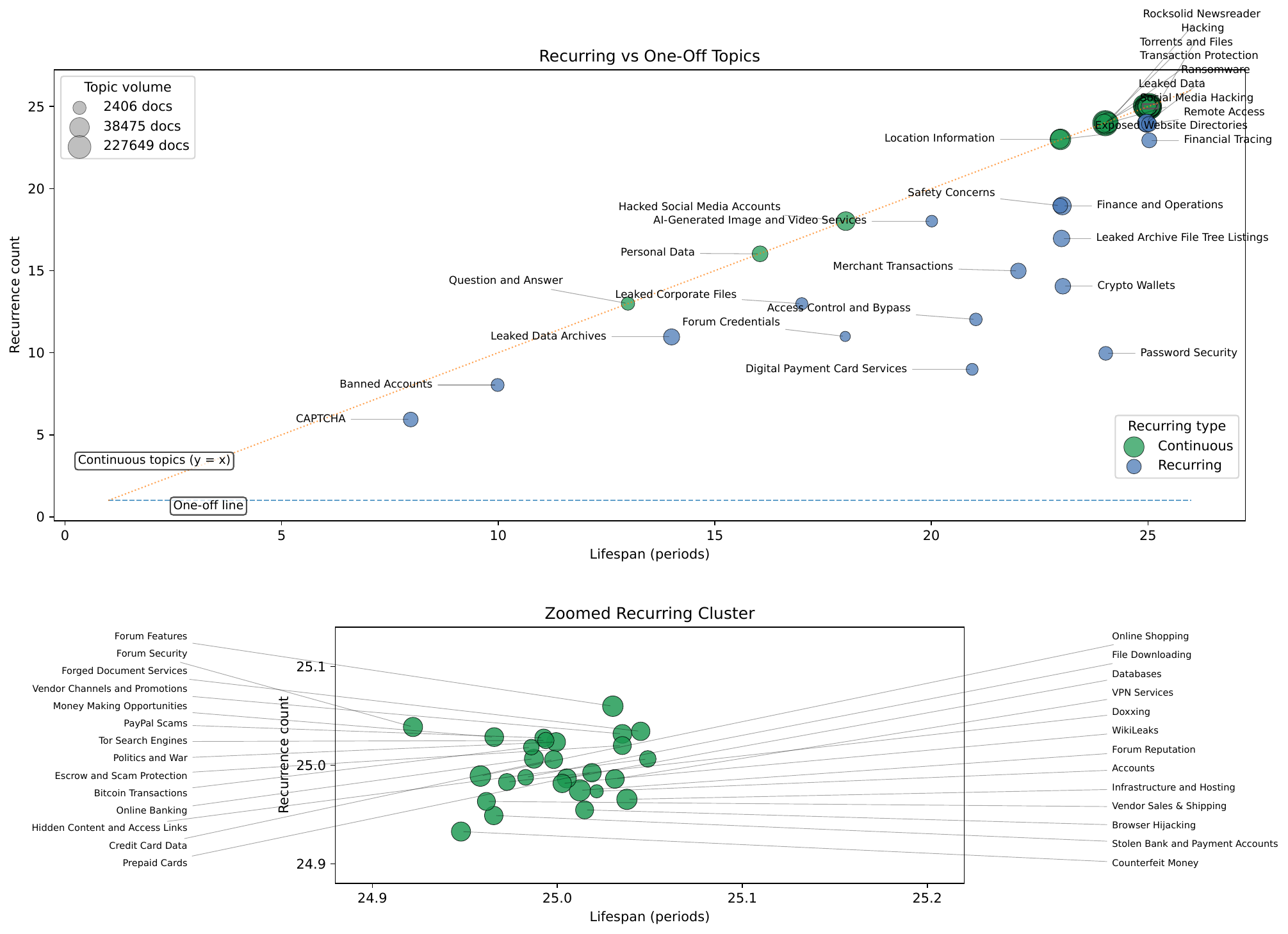}
    \caption{Recurring vs One-Off Topics.}
    \label{fig:rq2_recurring}
\end{figure}

% \subsubsection{Topic Lifespan Distribution}

% Topic lifespans varied widely across the dataset. Some topics appeared only 
% briefly, reflecting short-lived bursts of discussion activity, while others 
% persisted across large portions of the observation period.

% % Lifespan was measured as the time interval between the first and last observed appearance of a topic in the dataset. The resulting distribution 
% % reveals a mixture of transient and persistent discussion themes within the 
% % cybercrime ecosystem.

% % Histogram
% % X-axis: lifespan duration
% % Y-axis: number of topics

% \subsubsection{Growth and Decay Dynamics}

% Beyond lifespan, topics also exhibited different patterns of growth and 
% decline. Some topics emerged rapidly and disappeared shortly afterward, 
% indicating burst-like activity patterns. Others maintained relatively stable 
% levels of discussion over extended periods.

% These dynamics suggest that topic evolution typically occurs through gradual 
% changes in discussion volume rather than abrupt replacement of existing 
% themes.

% % Survival curve
% % X-axis: time
% % Y-axis: probability topic remains active

% % Normalized lifecycle line plot
% % X-axis: relative time to peak
% % Y-axis: normalized activity

\vspace{-3mm}
\subsubsection{Recurring vs One-Off Topics}

Figure~\ref{fig:rq2_recurring} shows a strong concentration of topics along the diagonal (recurrence $\approx$ lifespan), indicating that a large share of topics are active in nearly all observed periods. These continuous topics reach maximum lifespan (25 periods) with consistently high recurrence counts, suggesting persistent and sustained activity. In contrast, a smaller set of topics is located in the lower-left region of the figure, characterized by low lifespan ($<10$ periods) and low recurrence ($<10$), indicating short-lived and episodic behavior.

The distribution is therefore bimodal, with limited presence of topics in intermediate ranges of lifespan and recurrence. This suggests that topics tend to be either stable and continuously active or short-lived and sporadic, rather than gradually emerging and decaying. The zoomed cluster further shows that even among long-lived topics (lifespan $\approx 25$), recurrence varies substantially, reflecting differences in activity intensity rather than survival.

\vspace{-3mm}
\subsubsection{Quantitative lifecycle indicators} The lifecycle results provide a more explicit picture of topic persistence and change. The median topic lifespan is 75 months (25 periods), while the mean lifespan is 68.45 months (22.82 periods). The longest surviving topics include \textit{Forum Reputation}, \textit{Online Shopping}, and \textit{Forum Features}, whereas the shortest-lived topics include \textit{CAPTCHA}, \textit{Banned Accounts}, and \textit{Question and Answer}. These results confirm that most topics remain active for substantial portions of the observation window, while genuinely short-lived grouped themes are comparatively rare.

Growth and decline are similarly uneven but bounded. The strongest positive trends are observed for \textit{Forum Features}, \textit{Torrents and Files}, and \textit{Infrastructure and Hosting}, while the steepest negative trends are observed for \textit{Forum Reputation}, \textit{WikiLeaks}, and \textit{Online Shopping}. Among topics that disappear before the final observed period, the mean time from peak activity to last observed activity is 28 months. This indicates that even declining topics generally fade over multiple periods and not disappear immediately.

% THIS BELOW SHOULD BE IN THE DISCUSSION
% \subsection{Structural Roles of Topics}

% Combining prevalence and lifecycle measures provides additional insight into 
% the structural roles that topics play within the ecosystem.

% Topics that exhibit both high prevalence and long lifespans function as 
% \textit{core} activities within cybercrime communities, representing stable 
% areas of demand and ongoing discussion. In contrast, topics with low 
% prevalence and short lifespans represent more opportunistic or temporary 
% interests.

% This distinction reveals a structural pattern in which a small set of 
% persistent topics anchors the ecosystem, while numerous short-lived themes 
% appear intermittently around these core activities.

\begin{cooltextbox}
    \textsc{\textbf{Answer to RQ2:}}
    Topic lifecycles are dominated by persistence and gradual change. The median topic lifespan is 75 months, and even the shortest-lived grouped topics remain active for multiple years. Changes in thematic prominence occur mainly through gradual growth and decline rather than abrupt emergence or disappearance, indicating that cybercrime ecosystems evolve through slow redistribution of attention within a stable topic structure.
\end{cooltextbox}
\section{Discussion}
\label{sec:discussion}
\noindent

\noindent We discuss our key findings (§\ref{ssec:key_findings}), outline implications (§\ref{ssec:implications}) as well as limitations and future work (§\ref{ssec:limitations}).

\subsection{Key Findings} \label{ssec:key_findings}

\noindent The dark web ecosystem is both structurally concentrated and thematically persistent. A small number of topics account for most observed activity: the top five topics explain 53.15\% of discussion volume, the top ten explain 68.67\%, and the top twenty explain 88.36\% of the corpus. This concentration indicates that cybercrime ecosystems are not organized around a broad and continuously shifting set of dominant concerns, but instead around a limited set of stable and repeatedly observed functions.

This concentration is closely linked to their persistence. At the level of the final grouped topic structure, all identified topics are either recurring or continuous rather than one-off. In other words, the most important themes are not only prevalent but also durable over time. Thus, the ecosystem is shaped less by novelty than by continuity in economically relevant activities such as exchange, coordination, reputation building, infrastructure use and transaction protection.

The lifecycle analysis confirms this interpretation. Topics are long-lived, with a median lifespan of 75 months and a mean lifespan of 68.45 months. Even the shortest-lived grouped topics remain active for at least two years. This indicates that thematic change in cybercrime ecosystems rarely takes the form of abrupt emergence and disappearance. Instead, changes occur through gradual growth, decline and redistribution of attention within an already established thematic structure.

At the same time, the ecosystem is not static. Several topics show meaningful directional trends. \textit{Forum Features}, \textit{Torrents and Files}, and \textit{Infrastructure and Hosting} exhibit the strongest positive trajectories, while \textit{Forum Reputation}, \textit{WikiLeaks}, and \textit{Online Shopping} show the strongest declines. However, even among topics that disappear before the end of the observation period, the average lag between peak activity and disappearance is 28 months. This indicates that decline is typically progressive.

\subsection{Implications}
\label{ssec:implications}

\noindent Our findings highlight that cybercrime ecosystems evolve through stable, long-lived topics with gradual shifts in prominence, which has several implications for CTI:

\noindent \textbf{Longitudinal monitoring is essential.} State-of-the-art approaches miss whether topics are emerging, declining or stable. Tracking topic trajectories over time reveals meaningful dynamics that are otherwise invisible.

\noindent \textbf{Resource allocation should be stratified.} Persistent core topics (e.g., \textit{Online Shopping}, \textit{Infrastructure and Hosting}) need continuous automated monitoring, while emerging topics (e.g., \textit{Forum Features}, \textit{Databases}) require periodic expert analysis. Short-lived spikes can be handled via lightweight alerts. This allows more efficient use of analyst effort.

\noindent \textbf{Distinguishing signal from noise improves operational intelligence.} Persistent topics provide reliable intelligence sources, while sustained growth indicates structural change. In contrast, short-lived bursts often reflect temporary fluctuations and should not be overinterpreted.

\noindent \textbf{Longitudinal analysis enables evaluation of enforcement impact.} Changes in topic prevalence can indicate whether interventions lead to suppression, displacement or rapid recovery. The observed multi-year topic lifespans provide a baseline for distinguishing temporary disruption from structural change.

\noindent \textbf{Threat intelligence should adopt longer planning horizons.} Since most topics persist for years, organizations can move from reactive monitoring to multi-year strategic planning. Sustained trends justify long-term investment, whereas transient topics do not.

\subsection{Limitations and Future Work}
\label{ssec:limitations}
\noindent We outline limitations that should be considered when interpreting the results and reflect upon future work.

\noindent \textbf{Dataset Constraints}  As the dataset was collected by an external organization using proprietary crawling strategies, the coverage of the dark web ecosystem cannot be considered exhaustive. Snapshot frequency varies across websites and time periods, which may introduce temporal bias where increased crawling is misinterpreted as increased activity. Website identification based on file path and page title may not fully capture rebranding, domain changes, or migration, which can potentially underestimate continuity. Cross-platform activity (e.g., migration to Telegram) is not captured, so observed declines may reflect relocation rather than disappearance. Topic prevalence reflects content volume only, as no traffic or user engagement data are available. Additionally, the observation window (2020–2026) includes major external events but lacks earlier baseline data, which limits our comparison with longer-term trends.

\noindent \textbf{Modeling Assumptions} Our focus on English-language content ensures semantic consistency but excludes non-English communities, limiting generalizability to the global cybercrime ecosystem. Topic modeling parameters were fixed after initial evaluation. Although they produce coherent and interpretable topics, alternative configurations may reveal different levels of granularity or uncover niche activities currently treated as noise. Furthermore, topics are modeled as static entities over time, which does not capture semantic drift, splitting or merging of themes as discussions evolve.

%Our analysis focuses exclusively on English-language content. While this ensures semantic consistency in topic modeling, it excludes non-English discussions that may represent substantial cybercrime activity, particularly in Russian, Chinese, and other language communities. Our findings therefore reflect English-language cybercrime communities rather than the global dark web ecosystem. In the context of this constraint, we selected our topic modeling parameters through comparative evaluation and kept them fixed throughout analysis. While these settings produced semantically coherent topics, alternative configurations may reveal different granularities of thematic structure. In particular, smaller cluster sizes might detect niche activities that our configuration treats as noise, while larger clusters might merge distinct but related themes. Our approach also treats topics as static entities throughout the observation period. Dynamic topic models could better capture gradual semantic drift or thematic splitting and merging over time.

\noindent \textbf{Temporal Aggregation} Aggregating snapshots into quarterly intervals reduces noise and mitigates crawling artifacts, but may overlook short-term dynamics. Thus, we may miss events such as enforcement actions, marketplace shutdowns or sudden activity spikes.

%We aggregated snapshots into quarterly intervals to balance noise reduction and temporal resolution. This window captures persistent trends while filtering crawl-related artifacts, but may mask rapid responses to specific enforcement actions or platform disruptions that unfold over days or weeks. For example, immediate post-takedown migration patterns or panic-driven discussion spikes may be smoothed out in quarterly aggregation. Alternative temporal resolutions (e.g., monthly or yearly windows) however were not systematically evaluated. Finer granularity might reveal short-term volatility in response to specific events, while coarser intervals might better isolate long-term structural trends. The quarterly window represents our initial choice based on common practice in longitudinal studies, but its optimality for detecting different types of temporal dynamics remains untested. This limits conclusions about phenomena that operate at significantly faster or slower timescales than quarters.

\noindent \textbf{Platform Differentiation} Forums and marketplaces are modeled jointly to enable ecosystem-level analysis and cross-platform comparison. While this provides a unified view, it may obscure platform-specific practices, vocabulary and structural differences (e.g., in platform types).

\noindent \textbf{Temporal Validation} Our manual validation focused on topic coherence and label accuracy. While we ensured that observed trends reflect actual content changes rather than crawling artifacts, we did not systematically validate topic trajectories against external event timelines, such as enforcement actions or publicly reported data breaches. Interpretations of temporal patterns are therefore based on post-hoc contextualization. 

%Manual validation focused on topic coherence and label accuracy. While we verified that observed trends corresponded to actual content changes rather than crawling artifacts, we did not systematically validate specific topic trajectories against external event timelines, such as confirmed enforcement actions or publicized data breaches. In our discussion, we interpret temporal patterns in relation to known external events, but this represents post-hoc contextualization rather than prospective validation. Future work could strengthen temporal validity by systematically triangulating topic dynamics with documented external events before interpretation.

\vspace{0.3cm}
\noindent \textbf{Future Directions} We could address these limitations via several directions: (i) controlled crawling with consistent snapshot intervals would eliminate temporal bias from uneven crawling schedules and enable finer-grained analysis of short-term dynamics; (ii) extending the framework to non-English content would provide a more complete view of the global cybercrime ecosystem; (iii) separate topic modeling for forums, marketplaces, and other platform types could reveal specialized practices and vocabularies; (iv) incorporating temporal dependencies directly into the modeling process (e.g., using dynamic topic models or sequential clustering) could better capture how topics evolve, split or merge; (v) integrating clear web sources (e.g., Telegram channels, paste sites) would enable tracking of activity migration across infrastructures; (vi) systematic evaluation of clustering parameters could establish robustness bounds and identify whether key findings hold across alternative configurations; (vii) analyzing cross-topic correlations to capture dependencies between themes; and (viii) building on longitudinal topic trajectories to predict topic persistence, emergence or platform migration patterns could allow for forward-looking CTI threat assessment.
\section{Conclusions}
\label{sec:conclusions}

\noindent This study examines how content on the dark web evolves to identify patterns relevant to cybersecurity and CTI. Using NLP and topic modeling on longitudinal snapshots, we captured thematic development across a large and diverse set of sites. The results show that cybercrime ecosystems evolve primarily through gradual shifts in existing topics, while maintaining a stable underlying structure. 

%\textbf{RQ1} demonstrates that topic prevalence is highly concentrated, with a dominating small set of core topics. \textbf{RQ2} shows that topics persist over long periods and change through gradual growth and decline, with little evidence of abrupt emergence or disappearance. Thus, the change usually occurs through redistribution of attention within a fixed set of activities. 

\section*{ACKNOWLEDGMENT}
\noindent We thank CFLW Cyber Strategies for data collected with their Dark Web Monitor. Part of this research was funded by Hilti. This publication is part of the INTERSECT project, Grant No. NWA.1162.18.301, funded by NWO and the CATRIN project, Grant No. NWA.1215.18.003. %(for RR and LA)

\begin{comment}
\noindent This study examined how content and external links on Darknet websites evolve over time, with the aim of identifying patterns relevant to cybersecurity and CTI. Using NLP and topic modeling on longitudinal snapshots, the analysis captured both thematic development and linking behavior across a diverse set of Darknet sites. The findings indicate that cybercrime websites in the dark web evolve through gradual shifts in existing topics rather than abrupt thematic change, while link dynamics reflect a mix of stable and fragmented structures consistent with the decentralized nature of the ecosystem.

With respect to the research questions, \textbf{RQ1} shows that topic prevalence is structurally stable and highly concentrated, with a small set of core topics consistently dominating activity while changes occur as shifts in their relative importance. \textbf{RQ2} indicates that topics persist over long periods and evolve through gradual growth and decay, with limited evidence of abrupt introduction or disappearance.

Cybercrime forums and marketplaces on the dark web are stable but adaptive: change occurs through redistribution of attention within a fixed set of activities rather than through disruption. This highlights the importance of longitudinal analysis for CTI, as meaningful signals lie in sustained trends and gradual changes rather than common short-term fluctuations.
\end{comment}
%\input{sections/0-acknowledgements}

\bibliographystyle{IEEEtran}

{\footnotesize
\bibliography{bibliography} 
}
% \newpage

\appendices
\section{}

\subsection{Reproducibility Overview}
\label{app:reproducibility}

\noindent This section summarizes the configuration required to reproduce the content analysis pipeline. All preprocessing and modeling steps were executed offline on archived HTML snapshots using deterministic settings where applicable. The pipeline was implemented in Python using the following libraries:

\begin{itemize}
\item BERTopic
\item HuggingFace and sentence-transformers
\item UMAP-learn and HDBSCAN
\item scikit-learn
\item Trafilatura (HTML text extraction)
\end{itemize}

HTML content was extracted using Trafilatura's default extraction settings and only English-language text was retained. Text preprocessing included:

\begin{itemize}
\item lowercase normalization
\item removal of URLs, email addresses, punctuation, numbers, and special characters
\item WordNet lemmatization
\item standard English stopword removal
\item token length filtering (3--25 characters)
\end{itemize}

Websites with fewer than four snapshots were excluded from the dataset.

Semantic embeddings were generated using \textit{DarkBERT} sentence embeddings, selected through comparative evaluation (Appendix~\ref{app:parameter-evaluation}). Dimensionality reduction and clustering used the following configuration:

\begin{itemize}
\item \textbf{UMAP:} $n\_components=5$, metric=cosine, fixed random state
\item \textbf{HDBSCAN:} metric=euclidean, min\_cluster\_size= 80, min\_samples= 90, prediction\_data= True
\end{itemize}

Topic representation used class-based TF-IDF (c-TF-IDF) with KeyBERT keyphrase refinement. Topic labels were generated using the \textit{Llama} language model from representative keywords and sample documents, followed by manual validation (Appendix~\ref{app:human-validation}). Snapshots were assigned probabilistic topic distributions using BERTopic soft assignment.

For temporal analysis, snapshots were ordered chronologically at the website level and topic probabilities were aggregated within time intervals to compute the longitudinal prevalence and topic turnover measures used in the main analysis.

\subsection{Dataset Filtering and Temporal Handling}
\label{app:filtering and temporal}

\noindent This section documents the criteria used to determine which websites, snapshots, and textual content were included in the analysis. The goal of the filtering process was to ensure longitudinal consistency while minimizing noise unrelated to cybercrime discussions.

\subsubsection{Details on the Dataset}
\label{app:details_dataset}
The dataset consists of archived HTML snapshots of dark web forums and marketplaces collected between 2020 and 2026 by CFLW Cyber Strategies, a third-party provider specializing in cyber threat intelligence. It includes textual content from publicly accessible pages such as forum discussions, marketplace listings, and related informational pages. In line with our definition of cybercrime as economically motivated activity, the dataset was pre-filtered by the provider to exclude content categories outside this scope. This includes explicit adult material, purely ideological or political content (e.g., extremism), and other non-economic or non-cybercrime-related domains. Additionally, personally identifiable information and sensitive operational details were removed during preprocessing. The resulting dataset focuses on textual content relevant to economically driven cybercriminal activity, suitable for aggregate analysis of ecosystem-level patterns.

\subsubsection{Website Selection}

The dataset consists of archived HTML snapshots collected over multiple years. A website was defined as a stable page identity observed repeatedly across time. To construct longitudinal histories:

\begin{itemize}
    \item Snapshots were grouped by identical file path and page title
    \item Each group was assigned a unique internal website identifier
\end{itemize}

Websites with fewer than four snapshots were excluded, as they do not provide sufficient temporal coverage for trend analysis. The number of snapshots per website varies substantially (mean = 321.42, median = 10), reflecting differences in crawling frequency and site availability. As continuous uptime information is not available, website persistence is approximated through repeated observations across time.

\subsubsection{Snapshot Validity Criteria}

Individual snapshots were retained only if they satisfied all of the following conditions:

\begin{itemize}
    \item A valid timestamp was available
    \item HTML content could be successfully parsed
    \item Extracted text length exceeded 50 characters
\end{itemize}

Snapshots failing any condition were discarded to prevent noise from incomplete crawls or placeholder pages.

\subsubsection{Content-Based Filtering}

After text extraction, additional filtering steps were applied:

\textbf{Language Restriction.} Only English-language content was retained. Non-English snapshots were excluded to ensure semantic consistency in embedding and topic modeling.

\textbf{Duplicate Removal.} Near-duplicate snapshots within the same website and time interval were removed. Two snapshots were considered duplicates if their cleaned textual content matched after normalization.

\textbf{Non-Informational Pages.} Pages containing only navigation elements, login prompts, error pages, or empty marketplace listings were excluded. These pages typically lack meaningful discussion content and can bias topic distributions.

\subsubsection{Temporal Consistency Handling}

Snapshots were ordered chronologically using timestamps. When multiple snapshots existed within the same time interval, they were aggregated during analysis rather than removed, ensuring that content updates were preserved without overweighting high-frequency crawls.

The filtering criteria were designed to balance coverage and reliability:

\begin{itemize}
    \item Minimum snapshot requirement ensures longitudinal validity
    \item Language filtering ensures semantic comparability
    \item Content filtering removes structural website noise
\end{itemize}

These steps produce a dataset suitable for measuring thematic prevalence and turnover over time, while reducing artifacts caused by crawling behavior or non-discussion pages.

\subsection{Model Comparison Experiments}
\label{app:parameter-evaluation}

\noindent This section documents the experimental comparison of candidate embedding models and clustering configurations used in the topic modeling pipeline. The objective was to select the configuration that maximized semantic coherence, minimized outlier assignments, and produced interpretable topic clusters suitable for longitudinal analysis.

\subsubsection{Candidate Embedding Models}

The following embedding models were evaluated:

\begin{itemize}
    \item MiniLM (all-MiniLM-L6-v2)
    \item DarkBERT
    \item AttackBERT
\end{itemize}

Each model was integrated into the BERTopic framework using identical dimensionality reduction and clustering procedures to ensure comparability.

\subsubsection{Evaluation Metrics}

Model performance was evaluated using a combination of quantitative clustering metrics and qualitative interpretability assessment.

\begin{itemize}
    \item \textbf{Topic Coherence (c\_v)} — measures semantic similarity among top words in each topic. Higher values indicate more interpretable topics.
    \item \textbf{Number of Topics} — constrained to remain below 100 to maintain analytical usability while preserving granularity.
    \item \textbf{Outlier Count} — number of documents not assigned to any cluster. Higher values indicate stricter noise filtering but risk excluding meaningful content.
    \item \textbf{Cluster Stability} — consistency of topic structure across repeated runs with different random seeds.
    \item \textbf{Cluster Size Distribution} — minimum and average cluster size used to evaluate fragmentation vs overgeneralization.
    \item \textbf{Keyword Interpretability} — manual inspection of representative keywords and documents.
\end{itemize}

\subsubsection{Embedding Model Comparison}

We evaluated three embedding models (MiniLM, DarkBERT, and AttackBERT) under identical BERTopic configurations across three clustering strategies:

\begin{itemize}
    \item min\_cluster\_size = min\_samples
    \item min\_samples = min\_cluster\_size - 10
    \item min\_samples = min\_cluster\_size + 10
\end{itemize}

Evaluation considered topic granularity (75–90 topics target), outlier filtering, and minimum cluster population.

\autoref{tab:clustering_models_evaluation} report clustering outcomes under progressively stricter density constraints. The equal-density setting provides a baseline comparison, while decreasing min\_samples relaxes cluster acceptance and increasing it enforces stronger semantic separation. Across configurations, DarkBERT consistently produced topic counts within the desired analytical range and assigned more documents as outliers, indicating more selective clustering. This behavior suggests improved embedding geometry where unrelated discussions are rejected rather than weakly merged.

\begin{table}[ht]
\centering
\caption{Model Comparison Across Parameters (Best result in bold)}
\label{tab:clustering_models_evaluation}
\begin{tabular}{lcccc}
\hline
\textbf{Model} & \textbf{Parameters} & \textbf{Topics} & \textbf{Outliers} & \textbf{Min} \\
\hline

\multicolumn{5}{c}{\textbf{AttackBERT}} \\
\hline
AttackBERT & 500\_50   & 2392 & 1021396 & 500 \\
AttackBERT & 700\_70   & 1994 & 975141  & 701 \\
AttackBERT & 900\_90   & 1774 & 901760  & 900 \\
AttackBERT & 1200\_120 & 1532 & 975868  & 1200 \\
AttackBERT & 1500\_150 & 1274 & 963499  & 1500 \\
AttackBERT & 2000\_200 & 820  & 1235456 & 2001 \\
AttackBERT & 2500\_250 & 659  & 1391916 & 2510 \\
AttackBERT & 3000\_300 & 568  & 1507640 & 3004 \\
\hline
\multicolumn{5}{c}{\textbf{DarkBERT}} \\
\hline
DarkBERT & 500\_50   & 2323 & 1377521 & 501 \\
DarkBERT & 700\_70   & 2021 & 975145  & 700 \\
\textbf{DarkBERT} & \textbf{900\_90} & \textbf{1791} & \textbf{845592} & \textbf{900} \\
DarkBERT & 1200\_120 & 1596 & 855891  & 1203 \\
DarkBERT & 1500\_150 & 1407 & 838806  & 1500 \\
DarkBERT & 2000\_200 & 837  & 1173336 & 2008 \\
DarkBERT & 2500\_250 & 695  & 1286506 & 2503 \\
DarkBERT & 3000\_300 & 607  & 1375889 & 3000 \\

\hline
\multicolumn{5}{c}{\textbf{MiniLM}} \\
\hline
MiniLM & 500\_50   & 2272 & 1062998 & 500 \\
MiniLM & 700\_70   & 2006 & 1101910 & 700 \\
MiniLM & 900\_90   & 1785 & 1006363 & 900 \\
MiniLM & 1200\_120 & 1602 & 884327  & 1200 \\
MiniLM & 1500\_150 & 1468 & 937896  & 1504 \\
MiniLM & 2000\_200 & 921  & 1141357 & 2000 \\
MiniLM & 2500\_250 & 707  & 1377228 & 2506 \\
MiniLM & 3000\_300 & 629  & 1526143 & 3000 \\

\hline
\end{tabular}
\end{table}
\subsubsection{Results}

Qualitative inspection of resulting topics further confirmed that domain-specific embeddings produced more meaningful clusters. DarkBERT grouped infrastructure-related terms (e.g., tor, onion, server), marketplace terminology (e.g., product, paypal, western union), and community interaction terms (e.g., message, member, buyer protection) into coherent themes, whereas general-purpose embeddings produced clusters dominated by generic web vocabulary. These observations supported the final model selection.

\subsection{Labelling Topics with Llama}
\label{app:labelling_topics}

\noindent Llama was used for semantic structuring and annotation consistency rather than primary classification.

\subsubsection{Prompt Structure}

All prompts followed a structured format consisting of a task definition, labeling rules, and an explicit output constraint. The model was instructed to generate concise topic labels directly from representative keywords.

The exact prompt template used is shown below:

\begin{verbatim}
Task: Generate a short topic label.
Rules:
- Output ONLY the label
- Maximum 4 words
- No punctuation
- No explanation

Topic words: <top words>
Label:
\end{verbatim}

Here, \texttt{<top words>} represents the list of representative keywords extracted for each topic using c-TF-IDF and KeyBERT.

\subsubsection{Generation Settings}

Topic labels were generated using controlled sampling to balance consistency and flexibility. The generation parameters were:

\begin{itemize}
    \item Temperature: 0.1
    \item Sampling: enabled (\texttt{do\_sample=True})
    \item Maximum new tokens: 12
    \item End-of-sequence token: default model EOS token
\end{itemize}

A low temperature was used to ensure stable and reproducible outputs while still allowing minor variation when generating short labels. The combination of constrained prompting and limited token generation ensured that outputs remained concise and consistent across topics.

\subsection{Human Validation of Topic Labels}
\label{app:human-validation}

\noindent To ensure that generated topic labels accurately reflected the underlying keyword clusters, a manual validation step was performed. Topic labels were initially generated using the large language model \textit{Llama-3.1-8B-Instruct}, which produced candidate topic names based on representative keywords extracted for each cluster.

All generated labels were then manually reviewed. For each topic, the Llama-generated label was evaluated against the corresponding keyword set to determine whether it accurately represented the underlying theme of the cluster. Labels that were judged to be representative were retained without modification. When a label did not adequately describe the keyword set or produced an incorrect interpretation of the topic, a corrected label was assigned manually based on the keywords and their semantic context.

\subsubsection{Validation Procedure}

Two independent annotators reviewed all automatically generated topic labels produced by the topic modeling pipeline. Annotators assessed whether each label accurately represented the underlying documents and associated keywords, and performed corrections where necessary.

The validation process consisted of three main actions: (1) cleaning and rewording labels to improve clarity and remove non-descriptive or extraneous text, (2) aligning labels with domain-relevant terminology, and (3) identifying and merging semantically overlapping topics. Annotators performed these steps independently before meeting to resolve disagreements and produce a final, consensus-based set of topic labels. The inter-code reliability score was calculated following similar procedure presented in works by Pfister et al. \cite{pfister2025department} and Ave et al. \cite{ave2026botlane}: Cohen $\kappa$=0.96, which indicated strong agreeability.

\subsubsection{Types of Corrections}

The manual validation resulted in three primary types of refinements:

\textbf{Label Cleaning and Rewording.}
Many automatically generated labels contained extraneous instructions, formatting artifacts, or overly verbose descriptions (e.g., ``Here is the topic label: \ldots''). These were simplified into concise and interpretable labels.

\textbf{Semantic Correction.}
In some cases, labels did not accurately reflect the underlying keywords or documents. Annotators corrected these to better align with the actual topic content.

\textbf{Topic Merging.}
Multiple clusters were found to represent highly similar or identical concepts. Annotators identified such overlaps and merged them into a single, unified topic to avoid redundancy and improve analytical clarity.

\subsubsection{Examples of Label Corrections and Topic Merges}

Table~\ref{tab:validation_examples} presents representative examples of how topic labels were refined during manual validation.

\begin{table*}[htbp]
\centering
\caption{Examples of topic label corrections and merges after human validation}
\label{tab:validation_examples}
\begin{tabular}{p{2cm} p{4.5cm} p{4.5cm} p{4cm}}
\toprule
\textbf{Topic ID(s)} & \textbf{Llama Original Label(s)} & \textbf{Human Cleaned / Interpreted Label} & \textbf{Human Final Label} \\
\midrule

0 & Membership Status & Shop/Forum Membership & Forum Reputation \\

1, 31, 32, 38, 45, 52, 56, 57, 60, 62, 65, 70, 80 & 
Torrents and File Sizes; Torrent Files; Directories; Mobile Phone Torrents & 
Torrent File Listings / File Structures & 
Torrents and Files \\

19 & Transaction Support & Scam Warnings & Escrow and Scam Protection \\

15, 22, 25 & 
Transaction Protection; Vendor Protection; Cloned Credit Card Protection & 
Financial Protection Mechanisms & 
Transaction Protection \\

27 & 
Hacking and Password Security & 
Social Media Account Access & 
Social Media Hacking \\

12, 24, 82 & 
Counterfeit Currency and Surveillance; Counterfeit Money Transfer; Counterfeit Money & 
Counterfeit Financial Activity & 
Counterfeit Money \\

43, 44, 48, 49, 50, 59, 69, 71, 75 & 
Exposed Web Directories; File Listings; Directory Identifiers & 
Exposed File Systems & 
Exposed Website Directories \\

46, 53, 67 & 
Bitcoin; Bitcoin Transactions; Bitcoin Transaction & 
Bitcoin Payments & 
Bitcoin Transactions \\

\bottomrule
\end{tabular}
\end{table*}

\subsubsection{Discussion of Refinements}

The examples in Table~\ref{tab:validation_examples} illustrate common issues in automatically generated topic labels. First, prompt artifacts and non-descriptive instructions were frequently present in the original labels, requiring cleaning and reinterpretation. Second, several labels were overly broad or ambiguous, necessitating semantic correction based on the associated keywords. Finally, a substantial number of topics exhibited high overlap, particularly in domains such as file sharing, financial transactions, and leaked data, and were therefore merged to reduce redundancy.

\subsubsection{Impact on Final Topic Set}

The human validation process reduced noise, improved label clarity, and ensured that each topic corresponded to a distinct and meaningful concept. The resulting refined topic set forms the basis for all subsequent analyses of topic prevalence (RQ1) and lifecycle dynamics (RQ2).

\subsection{Detailed Topic Definitions}
\label{app:topics}

\noindent This appendix provides an overview of the topics identified in our analysis. Topics are grouped into higher-level categories reflecting their functional role within the ecosystem (e.g., transactional, product-related, infrastructure, and community-oriented).

% Topic labels were assigned through manual inspection of the most representative keywords and documents associated with each cluster. The short definitions provided in the tables are intended to clarify the primary focus of each topic and distinguish between closely related themes.

\begin{figure}[t]
    \centering
    \includegraphics[width=\columnwidth]{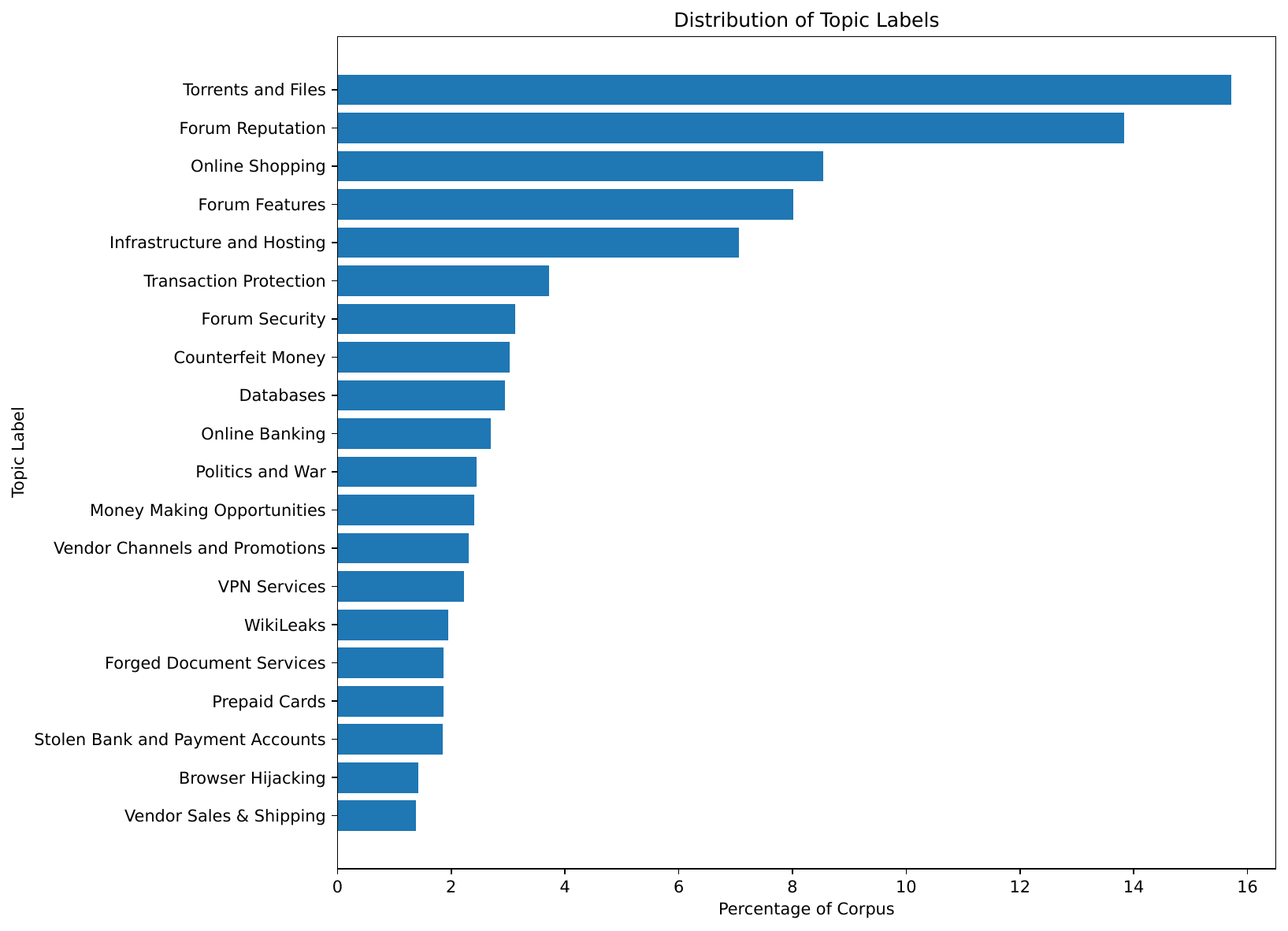}
    \caption{Distribution of topic labels in corpus.}
    \label{fig:darkbert_topic_distribution}
\end{figure}

For reference and clarity, each topic is assigned a unique identifier (e.g., T1, P3, I2, C4). Topics marked with $^{*}$ correspond to the ten most prevalent topics in the corpus, while those marked with $^{\dagger}$ fall within the top twenty. The distribution of the top 20 topic labels in the corpus is shown in \autoref{fig:darkbert_topic_distribution}.

\begin{table*}[t]
\centering
\caption{Transactional Topics}
\label{tab:topics_transactional}
\small
\begin{tabular}{p{1.2cm} p{4.3cm} r r p{6.5cm}}
\toprule
\textbf{ID} & \textbf{Topic} & \textbf{Count} & \textbf{\%} & \textbf{Description} \\
\midrule
T1 & Online Shopping$^{*}$ & 558{,}270 & 8.54 & Buying and selling goods or services. \\
T2 & Transaction Protection$^{\dagger}$ & 243{,}077 & 3.71 & Methods to secure or guarantee transactions. \\
T3 & Online Banking$^{\dagger}$ & 175{,}989 & 2.69 & Banking access, transfers, and fraud. \\
T4 & Money Making Opportunities$^{\dagger}$ & 156{,}609 & 2.39 & Methods for generating illicit income. \\
T5 & Prepaid Cards$^{\dagger}$ & 121{,}707 & 1.86 & Use and trade of prepaid payment cards. \\
T6 & Vendor Sales \& Shipping$^{\dagger}$ & 90{,}047 & 1.38 & Logistics and delivery of goods. \\
T7 & Escrow and Scam Protection$^{\dagger}$ & 89{,}166 & 1.36 & Third-party services to reduce fraud risk. \\
T8 & PayPal Scams$^{\dagger}$ & 74{,}718 & 1.14 & Fraud schemes involving PayPal. \\
T9 & Bitcoin Transactions & 22{,}191 & 0.34 & Payments using Bitcoin. \\
T10 & Crypto Wallets & 6{,}301 & 0.10 & Management of cryptocurrency wallets. \\
T11 & Merchant Transactions & 5{,}977 & 0.09 & Payment processing for vendors. \\
T12 & Financial Tracing & 4{,}778 & 0.07 & Tracking financial flows. \\
\bottomrule
\end{tabular}
\end{table*}

\begin{table*}[t]
\centering
\caption{Product-Related Topics}
\label{tab:topics_products}
\small
\begin{tabular}{p{1.2cm} p{4.3cm} r r p{6.5cm}}
\toprule
\textbf{ID} & \textbf{Topic} & \textbf{Count} & \textbf{\%} & \textbf{Description} \\
\midrule
P1 & Counterfeit Money$^{*}$ & 197{,}892 & 3.02 & Production and sale of fake currency. \\
P2 & Forged Document Services$^{\dagger}$ & 122{,}344 & 1.87 & Creation and sale of fake documents. \\
P3 & Stolen Bank and Pmt. Accs.$^{\dagger}$ & 120{,}963 & 1.85 & Compromised financial accounts. \\
P4 & Credit Card Data & 87{,}250 & 1.33 & Trade of stolen credit card data. \\
P5 & Social Media Hacking & 43{,}383 & 0.66 & Compromising social media accounts. \\
P6 & Hacked Social Media Accounts & 23{,}559 & 0.36 & Sale of hacked social accounts. \\
P7 & Ransomware & 16{,}101 & 0.25 & Malware used for extortion. \\
P8 & Hacking & 15{,}216 & 0.23 & General unauthorized access techniques. \\
P9 & Personal Data & 6{,}686 & 0.10 & Trade of personal information. \\
P10 & Accounts & 6{,}270 & 0.10 & Sale of compromised accounts. \\
\bottomrule
\end{tabular}
\end{table*}

\begin{table*}[t]
\centering
\caption{Infrastructure Topics}
\label{tab:topics_infrastructure}
\small
\begin{tabular}{p{1.2cm} p{4.3cm} r r p{6.5cm}}
\toprule
\textbf{ID} & \textbf{Topic} & \textbf{Count} & \textbf{\%} & \textbf{Description} \\
\midrule
I1 & Infrastructure and Hosting$^{*}$ & 461{,}151 & 7.05 & Hosting services and server infrastructure. \\
I2 & Databases$^{*}$ & 192{,}833 & 2.95 & Storage and access to data collections. \\
I3 & VPN Services$^{\dagger}$ & 145{,}070 & 2.22 & Privacy and anonymity via VPNs. \\
I4 & Browser Hijacking$^{\dagger}$ & 92{,}553 & 1.42 & Redirecting or controlling browsers. \\
I5 & Exposed Website Directories & 66{,}696 & 1.02 & Listings of accessible or vulnerable sites. \\
I6 & Hidden Cont. and Access Links & 38{,}594 & 0.59 & Access to restricted or hidden resources. \\
I7 & Rocksolid Newsreader & 38{,}475 & 0.59 & Tool for accessing content feeds. \\
I8 & Tor Search Engines & 33{,}629 & 0.51 & Search tools for dark web content. \\
I9 & Remote Access & 5{,}018 & 0.08 & Tools for remote system control. \\
I10 & CAPTCHA & 4{,}363 & 0.07 & CAPTCHA solving or bypass methods. \\
I11 & Access Control and Bypass & 2{,}015 & 0.03 & Circumventing access restrictions. \\
\bottomrule
\end{tabular}
\end{table*}

\begin{table*}[t]
\centering
\caption{Community Topics}
\label{tab:topics_community}
\small
\begin{tabular}{p{1.2cm} p{4.3cm} r r p{6.5cm}}
\toprule
\textbf{ID} & \textbf{Topic} & \textbf{Count} & \textbf{\%} & \textbf{Description} \\
\midrule
C1 & Torrents and Files$^{*}$ & 1{,}026{,}851 & 15.71 & File sharing via torrent systems. \\
C2 & Forum Reputation$^{*}$ & 903{,}883 & 13.83 & User trust, ratings, and credibility. \\
C3 & Forum Features$^{*}$ & 524{,}089 & 8.02 & Platform functionality and usability. \\
C4 & Forum Security$^{*}$ & 204{,}509 & 3.13 & Security practices within forums. \\
C5 & Vendor Channels and Promos.$^{\dagger}$ & 150{,}399 & 2.30 & Vendor advertising and communication. \\
C6 & WikiLeaks$^{\dagger}$ & 127{,}343 & 1.95 & Sharing and discussion of leaked information. \\
C7 & Politics and War$^{\dagger}$ & 159{,}594 & 2.44 & Discussions on geopolitical events. \\
C8 & Doxxing & 49{,}577 & 0.76 & Publishing personal information. \\
C9 & Location Information & 25{,}867 & 0.40 & Sharing geographic details. \\
C10 & File Downloading & 23{,}972 & 0.37 & General downloading practices. \\
C11 & Leaked Data & 19{,}650 & 0.30 & Discussion of leaked datasets. \\
C12 & Leaked Arch. File Tree Listings & 9{,}357 & 0.14 & Structured listings of leaked files. \\
C13 & Leaked Data Archives & 7{,}766 & 0.12 & Collections of leaked data. \\
C14 & Safety Concerns & 4{,}630 & 0.07 & Risk and precaution discussions. \\
C15 & Question and Answer & 2{,}594 & 0.04 & General help and information exchange. \\
C16 & Banned Accounts & 2{,}281 & 0.03 & Account restrictions and recovery. \\
C17 & Leaked Corporate Files & 1{,}732 & 0.03 & Corporate data breach discussions. \\
C18 & Forum Credentials & 977 & 0.01 & Access credentials for forum accounts. \\
C19 & Password Security & 3{,}014 & 0.05 & Password protection and cracking. \\
C20 & Finance and Operations & 16{,}096 & 0.25 & Operational coordination and planning. \\
C21 & Digital Payment Card Services & 1{,}605 & 0.02 & Services related to card payments. \\
\bottomrule
\end{tabular}
\end{table*}

\end{document}